# The Hitchhiker's Guide to the All-Interval 12-Tone Rows

*Marco Buongiorno Nardelli*

**Abstract**. This article revisits the generation, classification and categorization of all-intervals 12-tone series (AIS). Inspired by the seminal work of Morris and Starr in 1974 (Morris and Starr, The Structure of All-Interval Series 1974), it expands their analysis using complex network theory and provides composers and theorists with the re-ordering scheme that links all AISs together by chains of relations.

**Introduction**. Although all-interval 12-tone series (AIS) have been used in atonal music as early as the 1920's (A. Berg's *Lyric Suite* is an example), it took more than 40 years before their categorization and cataloguing. While Nicolas Slonimsky listed 18 "invertible dodecaphonic progressions with all different intervals" in his *Thesaurus of Scales and Melodic Patterns* (Slonimsky 1947), we owe Bauer-Mangelberg and Ferentz (Bauer-Mengelberg and Ferentz 1965) the first attempt to the generation of AIS using a computer program to explore all the combinatorial possibilities that give rise to such chords. Their approach led to a list of 1928 row "generators", AIS that produce other AIS by inversion or transposition. Their list was later purged by David Cohen in 1972 (Cohen 1972/73) to an "irreducible list" of 266 chords. The final and definitive list of AIS was published by Robert Morris and Daniel Starr only a couple of years later (Morris and Starr, The Structure of All-Interval Series 1974): using a more sophisticated computer program (and a more powerful computer!) they were able to identify the full corpus of the AIS: 3856 all-interval series that are transpositionally and rotationally normal (normal form), and that reduce to 1928 once inversionally related rows are eliminated. Once the AIS are identified computationally, the remaining of the discussion in their paper is based on a "by hand" analysis of the rows and their properties, and as such is limited by the power of human computation. In particular, they state that:

"It is possible that some […] re-ordering scheme could, possibly in conjunction with other operations, link all AISs together by chains of relations, which would be an elegant description of the generation of AISs and would doubtless provide interesting compositional devices for pitch-ordering." (Morris and Starr, The Structure of All-Interval Series 1974)

The above observation is the main motivation for this work: after introducing the computational tools used in our study, we indeed show that the "re-ordering" scheme envisioned by Morris and Starr exists in the form of complex networks and discuss the insight that can be gained by such representation.

**All-interval series generators**. We generate the full corpus of the AIS using an algorithm that explores all the permutations of the vector of the sequence of intervals in an ordered pitch class set (pcs) of cardinality 12. This vector, hereby referred to as LISV (linear interval sequence vector), can be interpreted as the step-interval vector introduced by Cohn (Cohn 1997) and originally proposed by Morris as the cyclic interval succession vector of stride 1 (Morris, Composition with pitch-classes: a theory of compositional design 1987). To produce the complete set of normal form generators of 12-tone rows, we construct all the possible permutations of the LISV vector that contains all the intervals and then reconstruct the rows starting with pc 0. Note that since the computation of the LISV exploits the cyclic property of the row, we need to end the interval series with a tritone, and then remove from the list all the rows that have cardinality less than 12. In Example 1 we show the few lines of python code needed to generate all the AIS in normal form.

```
perm = itertools.permutations([1,2,3,4,5,6,7,8,9,10,11])
last = 6
all_rows = []
for n in list(perm):
    ll = list(n)
    ll.append(last)
    row = [0]
    for j in ll:
        row.append((row[-1]+j)%12)
    row.pop()
    rown = Remove(row)
    if len(rown) == 12:
        all_rows.append(row)
```

Example 1. Python code that generates all the 3856 AIS in normal form. The `Remove` function removes all duplicate entries from the list, and we generate all the 11! permutations of the LISV using the `permutation` function in the python iterator module `itertools`.

Among the computational tools developed for this project (the IPython notebook with all the code to generate the results of this study will be made available as an on-line supplement to this paper), we have also written a helper class to seamlessly operate on the rows with the four closed symmetry operations of inversion (I), retrograde (R), element-wise multiplication of the series by 5, mod-12 (M) and the cyclic permutation of stride $w$ (Q), where $w$ is chosen so that the result is again an AIS in normal form (the results of Q and R need to be transposed to make them in normal form). This facilitates the classification of invariants and the identification of the irreducible normal forms, defined as "prime forms" from now on, as the subset of AIS that are not equivalent under I,R,M,Q (the subset of AIS that upon application of I,R,M,Q generate the full set of 3856 normal forms).

```
class AIrow:
#    Helper class for all-interval 12-tone rows operations (T,I,R,M,Q)
    def __init__(self,pcs,TET=12):
    def normalOrder(self):
    def intervals(self):
    def T(self,t=0):
    def I(self,pivot=0):
    def R(self):
    def Q(self):
    def M(self):
    def star(self):
    def constellation(self):
```

Example 2. Headers of the helper class `AIrow` methods: transposition (T), inversion (I), retrograde (R), multiplication (M), star and constellation (see text).

In Example 2 we show the headers of the helper class `AIrow` methods. Once we purge the list of all the rows that are related to each other by the four operations, we obtain the set of 918 prime form row generators, which we report in Table I and that will be analyzed in the remaining of this paper.

In Table 1 each prime form AIS is labelled with a sequential number (an extension of the Forte classification scheme (Forte 1973) for pitch class sets), and particular suffixes (S, P or L) are used to indicate some peculiar combinatorial property of the row discussed below. Of particular interest for the discussion that will follow, is the definition of "star" of the row: the set of rows that are produced by applying all the symmetry operations to a given AIS in prime form. It is closely related to what Morris and Starr call "constellation", the set of rows obtained by 16 simple and composite operations. Both the star and constellation for AIS 12-0P are shown in Example 3.

|   | Row |
|---|---|
| P | [ 0 1 3 7 2 5 11 10 8 4 9 6] |
| I | [ 0 11 9 5 10 7 1 2 4 8 3 6] |
| R | [ 0 3 10 2 4 5 11 8 1 9 7 6] |
| Q | [ 0 11 9 5 10 7 1 2 4 8 3 6] |
| M | [ 0 5 3 11 10 1 7 2 4 8 9 6] |

|    | P | I | IM | M |
|----|---|---|----|---|
| P  | [ 0 1 3 7 2 5 11 10 8 4 9 6] | [ 0 11 9 5 10 7 1 2 4 8 3 6] | [ 0 7 9 1 2 11 5 10 8 4 3 6] | [ 0 5 3 11 10 1 7 2 4 8 9 6] |
| R  | [ 0 3 10 2 4 5 11 8 1 9 7 6] | [ 0 9 2 10 8 7 1 4 11 3 5 6] | [ 0 9 10 2 4 11 5 8 7 3 1 6] | [ 0 3 2 10 8 1 7 4 5 9 11 6] |
| QR | [ 0 9 2 10 8 7 1 4 11 3 5 6] | [ 0 3 10 2 4 5 11 8 1 9 7 6] | [ 0 3 2 10 8 1 7 4 5 9 11 6] | [ 0 9 10 2 4 11 5 8 7 3 1 6] |
| Q  | [ 0 11 9 5 10 7 1 2 4 8 3 6] | [ 0 1 3 7 2 5 11 10 8 4 9 6] | [ 0 5 3 11 10 1 7 2 4 8 9 6] | [ 0 7 9 1 2 11 5 10 8 4 3 6] |

Example 3. Star (top) and constellation (bottom) of AIS 12-0P.

Going back to Table I, using the helper class `AIrow` it is easy to verify the invariant and combinatorial properties of the AIS. In particular, following Elliott Carter (Carter 2002), we identified 57 symmetrical inverted (non-retrogradable) and 34 parallel inverted AIS, labeled in Table 1 with the S and P suffixes, respectively.[1] Finally, the suffix L indicates the 121 prime form AIS that are Link chords, all-interval twelve-note chords, each of which contains either one or two instances of the all-trichord hexachord [0,1,2,4,7,8] as a contiguous subset. (Carter 2002)

---

[1] The symmetrically inverted chords correspond to the hexachordal combinatoriality where B=R(A), while the parallel inverted correspond to B=I(A).

| Label | Row | Intervals |
|---|---|---|
| 12-0P | [0, 1, 3, 7, 2, 5, 11, 10, 8, 4, 9, 6] | [ 1 2 4 7 3 6 11 10 8 5 9 6] |
| 12-1 | [0, 1, 3, 7, 2, 10, 8, 11, 5, 4, 9, 6] | [ 1 2 4 7 8 10 3 6 11 5 9 6] |
| 12-2 | [0, 1, 3, 7, 2, 11, 4, 10, 9, 5, 8, 6] | [ 1 2 4 7 9 5 6 11 8 3 10 6] |
| 12-3 | [0, 1, 3, 7, 2, 11, 10, 4, 9, 5, 8, 6] | [ 1 2 4 7 9 11 6 5 8 3 10 6] |
| 12-4 | [0, 1, 3, 7, 4, 11, 9, 8, 2, 5, 10, 6] | [ 1 2 4 9 7 10 11 6 3 5 8 6] |
| 12-5 | [0, 1, 3, 7, 4, 2, 9, 8, 11, 5, 10, 6] | [ 1 2 4 9 10 7 11 3 6 5 8 6] |
| 12-6 | [0, 1, 3, 7, 5, 8, 4, 10, 9, 2, 11, 6] | [ 1 2 4 10 3 8 6 11 5 9 7 6] |
| 12-7 | [0, 1, 3, 7, 5, 2, 9, 8, 11, 4, 10, 6] | [ 1 2 4 10 9 7 11 3 5 6 7 6] |
| 12-8 | [0, 1, 3, 7, 5, 2, 10, 4, 9, 8, 11, 6] | [ 1 2 4 10 9 8 6 5 11 3 7 6] |
| 12-9 | [0, 1, 3, 8, 11, 5, 9, 4, 2, 10, 7, 6] | [ 1 2 5 3 6 4 7 10 8 9 11 6] |
| 12-10 | [0, 1, 3, 8, 11, 7, 5, 4, 10, 2, 9, 6] | [ 1 2 5 3 8 10 11 6 4 7 9 6] |
| 12-11 | [0, 1, 3, 8, 11, 9, 5, 4, 10, 7, 2, 6] | [ 1 2 5 3 10 8 11 6 9 7 4 6] |
| 12-12L | [0, 1, 3, 8, 2, 5, 9, 7, 4, 11, 10, 6] | [ 1 2 5 6 3 4 10 9 7 11 8 6] |
| 12-13 | [0, 1, 3, 8, 2, 10, 5, 4, 7, 11, 9, 6] | [ 1 2 5 6 8 7 11 3 4 10 9 6] |
| 12-14 | [0, 1, 3, 8, 4, 7, 5, 11, 10, 2, 9, 6] | [ 1 2 5 8 3 10 6 11 4 7 9 6] |
| 12-15 | [0, 1, 3, 8, 5, 11, 7, 10, 9, 4, 2, 6] | [ 1 2 5 9 6 8 3 11 7 10 4 6] |
| 12-16 | [0, 1, 3, 8, 7, 10, 4, 11, 9, 5, 2, 6] | [ 1 2 5 11 3 6 7 10 8 9 4 6] |
| 12-17 | [0, 1, 3, 9, 2, 10, 5, 4, 7, 11, 8, 6] | [ 1 2 6 5 8 7 11 3 4 9 10 6] |
| 12-18 | [0, 1, 3, 9, 2, 10, 7, 5, 4, 8, 11, 6] | [ 1 2 6 5 8 9 10 11 4 3 7 6] |
| 12-19 | [0, 1, 3, 9, 2, 10, 8, 5, 4, 7, 11, 6] | [ 1 2 6 5 8 10 11 3 4 7 11 6] |
| 12-20 | [0, 1, 3, 9, 2, 11, 7, 10, 5, 4, 8, 6] | [ 1 2 6 5 9 8 3 7 11 4 10 6] |
| 12-21 | [0, 1, 3, 9, 5, 10, 8, 7, 4, 11, 2, 6] | [ 1 2 6 8 5 10 11 9 7 3 4 6] |
| 12-22 | [0, 1, 3, 9, 8, 4, 7, 5, 10, 2, 11, 6] | [ 1 2 6 11 8 3 10 5 4 9 7 6] |
| 12-23 | [0, 1, 3, 9, 8, 4, 2, 5, 10, 7, 11, 6] | [ 1 2 6 11 10 3 5 9 7 4 11 6] |
| 12-24 | [0, 1, 3, 10, 2, 5, 11, 7, 4, 9, 8, 6] | [ 1 2 7 4 3 6 8 9 5 11 10 6] |
| 12-25S | [0, 1, 3, 10, 2, 5, 11, 8, 4, 9, 7, 6] | [ 1 2 7 4 3 6 9 8 5 10 11 6] |
| 12-26 | [0, 1, 3, 10, 2, 8, 11, 7, 5, 4, 9, 6] | [ 1 2 7 4 6 3 8 10 11 5 9 6] |
| 12-27S | [0, 1, 3, 10, 2, 11, 5, 8, 4, 9, 7, 6] | [ 1 2 7 4 9 6 3 8 5 10 11 6] |
| 12-28 | [0, 1, 3, 10, 4, 8, 11, 9, 5, 2, 7, 6] | [ 1 2 7 6 4 3 10 8 9 5 11 6] |
| 12-29 | [0, 1, 3, 10, 4, 9, 8, 11, 7, 5, 2, 6] | [ 1 2 7 6 5 11 3 8 10 9 4 6] |
| 12-30 | [0, 1, 3, 10, 8, 7, 4, 9, 5, 11, 2, 6] | [ 1 2 7 10 11 9 5 8 6 3 4 6] |
| 12-31 | [0, 1, 3, 10, 9, 2, 5, 11, 7, 4, 8, 6] | [ 1 2 7 11 5 3 6 8 9 7 4 6] |
| 12-32 | [0, 1, 3, 10, 9, 5, 11, 2, 7, 4, 8, 6] | [ 1 2 7 11 8 6 3 5 9 4 10 6] |
| 12-33 | [0, 1, 3, 11, 2, 7, 4, 10, 5, 9, 8, 6] | [ 1 2 8 3 5 9 6 7 4 11 10 6] |
| 12-34 | [0, 1, 3, 11, 4, 10, 7, 2, 5, 9, 8, 6] | [ 1 2 8 5 6 9 7 3 4 11 10 6] |
| 12-35 | [0, 1, 3, 11, 4, 10, 8, 7, 2, 5, 9, 6] | [ 1 2 8 5 6 10 11 7 3 4 9 6] |
| 12-36 | [0, 1, 3, 11, 4, 2, 8, 7, 10, 5, 9, 6] | [ 1 2 8 5 10 6 11 3 7 4 9 6] |
| 12-37 | [0, 1, 3, 11, 5, 2, 7, 10, 9, 4, 8, 6] | [ 1 2 8 6 9 5 3 11 7 4 10 6] |
| 12-38 | [0, 1, 3, 11, 9, 4, 8, 2, 5, 10, 7, 6] | [ 1 2 8 10 7 4 6 3 5 9 11 6] |
| 12-39 | [0, 1, 3, 11, 9, 8, 5, 10, 4, 7, 2, 6] | [ 1 2 8 10 11 9 5 6 3 7 4 6] |
| 12-40 | [0, 1, 3, 2, 7, 10, 8, 4, 11, 5, 9, 6] | [ 1 2 11 5 3 10 8 7 6 4 9 6] |
| 12-41 | [0, 1, 3, 2, 9, 5, 10, 4, 7, 11, 8, 6] | [ 1 2 11 7 8 5 6 3 4 9 10 6] |
| 12-42 | [0, 1, 3, 2, 9, 7, 4, 8, 11, 5, 10, 6] | [ 1 2 11 7 10 9 4 3 6 5 8 6] |
| 12-43 | [0, 1, 4, 8, 10, 5, 3, 9, 2, 11, 7, 6] | [ 1 3 4 2 7 10 6 5 9 8 11 6] |
| 12-44S | [0, 1, 4, 8, 3, 5, 11, 9, 2, 10, 7, 6] | [ 1 3 4 7 2 6 10 5 8 9 11 6] |
| 12-45 | [0, 1, 4, 8, 7, 9, 2, 11, 5, 3, 10, 6] | [ 1 3 4 11 2 5 9 6 10 7 8 6] |
| 12-46L | [0, 1, 4, 8, 7, 2, 10, 3, 5, 11, 9, 6] | [ 1 3 4 11 5 8 5 2 6 10 6] |
| 12-47L | [0, 1, 4, 8, 7, 2, 11, 9, 3, 5, 10, 6] | [ 1 3 4 11 7 9 10 6 2 5 8 6] |
| 12-48 | [0, 1, 4, 8, 7, 5, 2, 10, 3, 9, 11, 6] | [ 1 3 4 11 10 9 8 5 6 2 7 6] |
| 12-49 | [0, 1, 4, 9, 11, 5, 2, 10, 8, 3, 7, 6] | [ 1 3 5 2 6 9 8 10 7 4 11 6] |
| 12-50 | [0, 1, 4, 9, 11, 8, 2, 10, 5, 3, 7, 6] | [ 1 3 5 2 9 8 7 10 4 11 6] |
| 12-51 | [0, 1, 4, 9, 3, 5, 2, 10, 8, 7, 11, 6] | [ 1 3 5 6 2 9 8 10 11 4 7 6] |
| 12-52 | [0, 1, 4, 9, 3, 11, 10, 8, 5, 7, 2, 6] | [ 1 3 5 6 8 11 10 9 2 7 4 6] |
| 12-53 | [0, 1, 4, 9, 3, 2, 10, 5, 7, 11, 8, 6] | [ 1 3 5 6 11 8 7 2 4 9 10 6] |
| 12-54 | [0, 1, 4, 9, 3, 2, 10, 8, 5, 7, 11, 6] | [ 1 3 5 6 11 8 10 9 2 4 7 6] |
| 12-55 | [0, 1, 4, 9, 5, 3, 2, 8, 10, 7, 11, 6] | [ 1 3 5 8 10 11 6 2 9 4 7 6] |
| 12-56 | [0, 1, 4, 10, 2, 7, 9, 8, 5, 3, 11, 6] | [ 1 3 6 4 5 2 11 9 10 8 7 6] |
| 12-57 | [0, 1, 4, 10, 3, 11, 9, 8, 5, 7, 2, 6] | [ 1 3 6 5 8 10 11 9 2 7 4 6] |
| 12-58 | [0, 1, 4, 10, 3, 2, 9, 8, 5, 7, 11, 6] | [ 1 3 6 5 11 7 8 2 4 9 10 6] |
| 12-59L | [0, 1, 4, 10, 5, 3, 11, 8, 7, 9, 2, 6] | [ 1 3 6 7 10 8 9 11 2 5 4 6] |
| 12-60 | [0, 1, 4, 10, 8, 3, 5, 9, 2, 11, 7, 6] | [ 1 3 6 10 7 2 4 5 9 8 11 6] |
| 12-61 | [0, 1, 4, 10, 8, 7, 3, 5, 9, 2, 11, 6] | [ 1 3 6 10 11 8 2 4 5 9 7 6] |
| 12-62 | [0, 1, 4, 10, 9, 5, 7, 2, 11, 3, 8, 6] | [ 1 3 6 11 8 2 7 9 4 5 10 6] |
| 12-63 | [0, 1, 4, 11, 3, 9, 2, 10, 8, 5, 7, 6] | [ 1 3 7 4 6 5 8 10 9 2 11 6] |
| 12-64 | [0, 1, 4, 11, 5, 2, 10, 3, 7, 9, 8, 6] | [ 1 3 7 6 9 8 5 4 2 11 10 6] |
| 12-65 | [0, 1, 4, 11, 9, 5, 2, 8, 10, 3, 7, 6] | [ 1 3 7 10 8 9 6 2 5 4 11 6] |
| 12-66 | [0, 1, 4, 11, 10, 3, 7, 9, 5, 2, 8, 6] | [ 1 3 7 11 5 4 2 8 9 6 10 6] |
| 12-67 | [0, 1, 4, 11, 10, 7, 9, 5, 3, 8, 2, 6] | [ 1 3 7 11 9 2 8 10 5 6 4 6] |
| 12-68L | [0, 1, 4, 11, 10, 7, 3, 5, 9, 2, 8, 6] | [ 1 3 7 11 9 8 2 4 5 6 10 6] |
| 12-69 | [0, 1, 4, 11, 10, 8, 2, 7, 3, 5, 9, 6] | [ 1 3 7 11 10 6 5 8 2 4 9 6] |
| 12-70 | [0, 1, 4, 11, 10, 8, 5, 7, 3, 9, 2, 6] | [ 1 3 7 11 10 9 2 8 6 5 4 6] |
| 12-71 | [0, 1, 4, 2, 7, 9, 8, 5, 11, 3, 10, 6] | [ 1 3 10 5 2 11 9 6 4 7 8 6] |
| 12-72 | [0, 1, 4, 2, 9, 3, 8, 5, 7, 11, 10, 6] | [ 1 3 10 7 6 5 9 2 4 11 8 6] |
| 12-73S | [0, 1, 4, 2, 9, 5, 11, 3, 8, 10, 7, 6] | [ 1 3 10 7 8 6 4 5 2 9 11 6] |
| 12-74 | [0, 1, 4, 2, 9, 5, 11, 8, 10, 3, 7, 6] | [ 1 3 10 7 8 6 9 2 5 4 11 6] |
| 12-75 | [0, 1, 4, 2, 10, 3, 7, 9, 8, 5, 11, 6] | [ 1 3 10 8 5 4 2 11 9 6 7 6] |
| 12-76P | [0, 1, 4, 2, 10, 3, 9, 8, 5, 7, 11, 6] | [ 1 3 10 8 5 6 11 9 2 4 7 6] |
| 12-77 | [0, 1, 4, 2, 10, 5, 11, 3, 8, 7, 9, 6] | [ 1 3 10 8 7 6 4 5 11 2 9 6] |
| 12-78 | [0, 1, 4, 2, 10, 9, 3, 8, 5, 7, 11, 6] | [ 1 3 10 8 11 6 5 9 2 4 7 6] |
| 12-79 | [0, 1, 4, 3, 7, 9, 2, 10, 5, 11, 8, 6] | [ 1 3 11 4 2 5 8 7 6 9 10 6] |
| 12-80 | [0, 1, 4, 3, 7, 9, 2, 10, 8, 5, 11, 6] | [ 1 3 11 4 2 5 8 10 9 6 7 6] |
| 12-81 | [0, 1, 4, 3, 7, 9, 5, 10, 8, 2, 11, 6] | [ 1 3 11 4 2 8 5 10 6 9 7 6] |
| 12-82 | [0, 1, 4, 3, 8, 2, 10, 5, 7, 11, 9, 6] | [ 1 3 11 5 6 8 7 2 4 10 9 6] |
| 12-83 | [0, 1, 4, 3, 8, 2, 10, 7, 5, 9, 11, 6] | [ 1 3 11 5 6 8 9 10 4 2 7 6] |
| 12-84 | [0, 1, 4, 3, 9, 2, 10, 5, 7, 11, 8, 6] | [ 1 3 11 6 5 8 7 2 4 9 10 6] |
| 12-85 | [0, 1, 4, 3, 9, 2, 10, 8, 5, 7, 11, 6] | [ 1 3 11 6 5 8 10 9 2 4 7 6] |
| 12-86 | [0, 1, 4, 3, 10, 2, 7, 9, 5, 11, 8, 6] | [ 1 3 11 7 4 5 2 8 6 9 10 6] |
| 12-87 | [0, 1, 4, 3, 10, 8, 5, 11, 7, 9, 2, 6] | [ 1 3 11 7 10 9 6 8 2 5 4 6] |
| 12-88P | [0, 1, 5, 7, 10, 3, 9, 8, 4, 2, 11, 6] | [ 1 4 2 3 5 6 11 8 10 9 7 6] |
| 12-89 | [0, 1, 5, 7, 10, 3, 2, 8, 4, 11, 9, 6] | [ 1 4 2 3 5 11 6 8 7 10 9 6] |
| 12-90 | [0, 1, 5, 7, 10, 8, 4, 9, 3, 2, 11, 6] | [ 1 4 2 3 10 8 5 6 11 9 7 6] |
| 12-91 | [0, 1, 5, 7, 10, 8, 4, 9, 3, 11, 2, 6] | [ 1 4 2 3 10 8 11 6 5 9 7 6] |
| 12-92 | [0, 1, 5, 7, 10, 9, 3, 8, 4, 2, 11, 6] | [ 1 4 2 3 11 6 5 8 10 9 7 6] |
| 12-93 | [0, 1, 5, 7, 2, 10, 4, 3, 8, 11, 9, 6] | [ 1 4 2 7 8 6 11 5 3 10 9 6] |
| 12-94 | [0, 1, 5, 7, 2, 10, 8, 11, 4, 3, 9, 6] | [ 1 4 2 7 8 10 3 5 11 6 9 6] |
| 12-95 | [0, 1, 5, 7, 2, 11, 9, 8, 4, 10, 3, 6] | [ 1 4 2 7 9 10 11 8 6 5 3 6] |
| 12-96 | [0, 1, 5, 7, 2, 11, 10, 8, 4, 9, 3, 6] | [ 1 4 2 7 9 11 10 8 5 6 3 6] |
| 12-97 | [0, 1, 5, 7, 3, 8, 11, 10, 4, 2, 9, 6] | [ 1 4 2 8 5 3 11 6 10 7 9 6] |
| 12-98P | [0, 1, 5, 7, 4, 9, 3, 2, 10, 8, 11, 6] | [ 1 4 2 9 5 6 11 8 10 3 7 6] |
| 12-99L | [0, 1, 5, 7, 4, 9, 3, 11, 2, 10, 8, 6] | [ 1 4 2 9 6 5 8 3 11 10 6] |
| 12-100 | [0, 1, 5, 7, 4, 11, 2, 10, 3, 9, 8, 6] | [ 1 4 2 9 7 3 8 5 6 11 10 6] |
| 12-101 | [0, 1, 5, 7, 4, 11, 2, 10, 9, 3, 8, 6] | [ 1 4 2 9 7 3 8 11 6 5 10 6] |
| 12-102 | [0, 1, 5, 7, 4, 11, 9, 8, 2, 10, 3, 6] | [ 1 4 2 9 7 10 11 6 8 5 3 6] |
| 12-103 | [0, 1, 5, 7, 2, 3, 9, 8, 11, 10, 4, 6] | [ 1 4 2 10 7 6 5 3 11 8 6 6] |
| 12-104 | [0, 1, 5, 7, 4, 2, 10, 3, 9, 11, 8, 6] | [ 1 4 2 9 10 8 5 6 11 3 7 6] |
| 12-105 | [0, 1, 5, 7, 4, 2, 10, 9, 3, 8, 11, 6] | [ 1 4 2 9 10 8 11 6 5 3 7 6] |
| 12-106 | [0, 1, 5, 7, 4, 3, 9, 2, 10, 8, 11, 6] | [ 1 4 2 9 11 6 5 8 10 3 7 6] |
| 12-107S | [0, 1, 5, 8, 10, 3, 9, 4, 2, 11, 7, 6] | [ 1 4 3 2 5 6 7 10 9 3 11 6] |
| 12-108 | [0, 1, 5, 8, 10, 4, 9, 7, 3, 2, 11, 6] | [ 1 4 3 2 6 5 10 8 11 9 7 6] |
| 12-109 | [0, 1, 5, 8, 3, 11, 10, 4, 2, 7, 9, 6] | [ 1 4 3 7 8 11 6 10 5 2 9 6] |
| 12-110 | [0, 1, 5, 8, 7, 2, 4, 10, 3, 11, 9, 6] | [ 1 4 3 11 7 2 6 5 8 10 9 6] |
| 12-111 | [0, 1, 5, 8, 7, 4, 2, 10, 3, 11, 9, 6] | [ 1 4 3 11 9 10 5 2 9 6] |
| 12-112L | [0, 1, 5, 10, 4, 7, 3, 2, 9, 11, 8, 6] | [ 1 4 5 6 3 8 11 7 2 9 10 6] |
| 12-113L | [0, 1, 5, 10, 4, 2, 9, 11, 8, 7, 3, 6] | [ 1 4 5 6 10 7 2 9 11 8 3 6] |
| 12-114L | [0, 1, 5, 10, 4, 2, 11, 7, 9, 8, 3, 6] | [ 1 4 5 6 10 9 8 2 11 7 3 6] |
| 12-115 | [0, 1, 5, 10, 7, 9, 4, 3, 11, 2, 8, 6] | [ 1 4 5 9 2 7 11 8 3 6 10 6] |
| 12-116L | [0, 1, 5, 10, 7, 3, 2, 8, 11, 9, 4, 6] | [ 1 4 5 9 8 11 6 3 10 7 2 6] |
| 12-117 | [0, 1, 5, 10, 8, 11, 7, 2, 4, 3, 9, 6] | [ 1 4 5 10 3 8 7 2 11 6 9 6] |
| 12-118 | [0, 1, 5, 10, 8, 2, 9, 11, 7, 4, 3, 6] | [ 1 4 5 10 6 7 2 8 9 11 6] |
| 12-119 | [0, 1, 5, 10, 8, 2, 11, 7, 9, 4, 3, 6] | [ 1 4 5 10 6 9 8 2 7 11 3 6] |
| 12-120 | [0, 1, 5, 10, 8, 4, 7, 9, 3, 2, 11, 6] | [ 1 4 5 10 8 3 2 6 11 9 7 6] |
| 12-121 | [0, 1, 5, 10, 9, 11, 7, 4, 2, 8, 3, 6] | [ 1 4 5 11 2 8 9 10 6 7 3 6] |
| 12-122 | [0, 1, 5, 11, 2, 9, 7, 4, 3, 8, 10, 6] | [ 1 4 6 3 7 10 9 11 5 2 8 6] |
| 12-123 | [0, 1, 5, 11, 2, 10, 7, 9, 3, 8, 4, 6] | [ 1 4 6 3 8 9 2 7 11 5 10 6] |
| 12-124 | [0, 1, 5, 11, 4, 2, 9, 8, 10, 7, 3, 6] | [ 1 4 6 5 10 7 11 2 9 8 3 6] |
| 12-125 | [0, 1, 5, 11, 4, 2, 10, 7, 9, 8, 3, 6] | [ 1 4 6 5 10 8 9 2 11 7 3 6] |
| 12-126 | [0, 1, 5, 11, 7, 4, 8, 3, 2, 4, 9, 6] | [ 1 4 6 8 3 10 7 8 5 2 9 6] |
| 12-127 | [0, 1, 5, 11, 7, 4, 2, 9, 8, 10, 3, 6] | [ 1 4 6 8 9 10 7 11 2 5 9 6] |
| 12-128 | [0, 1, 5, 11, 8, 10, 9, 4, 2, 7, 3, 6] | [ 1 4 6 9 2 11 7 10 5 8 3 6] |
| 12-129 | [0, 1, 5, 11, 8, 3, 2, 4, 9, 7, 10, 6] | [ 1 4 6 9 7 11 2 5 10 3 8 6] |
| 12-130L | [0, 1, 5, 11, 8, 7, 9, 4, 2, 10, 3, 6] | [ 1 4 6 9 11 2 7 10 8 5 3 6] |
| 12-131S | [0, 1, 5, 2, 4, 9, 3, 10, 8, 11, 7, 6] | [ 1 4 9 2 5 6 7 10 3 8 11 6] |
| 12-132 | [0, 1, 5, 2, 4, 10, 9, 7, 3, 8, 11, 6] | [ 1 4 9 2 6 11 10 8 5 3 7 6] |
| 12-133 | [0, 1, 5, 2, 4, 11, 7, 10, 3, 9, 8, 6] | [ 1 4 9 2 7 8 3 5 6 11 10 6] |
| 12-134 | [0, 1, 5, 2, 4, 11, 9, 7, 3, 3, 8, 6] | [ 1 4 9 2 7 8 3 11 6 10 6] |
| 12-135 | [0, 1, 5, 2, 4, 11, 9, 3, 8, 7, 10, 6] | [ 1 4 9 2 7 10 6 5 11 3 8 6] |
| 12-136 | [0, 1, 5, 2, 7, 10, 8, 4, 3, 9, 11, 6] | [ 1 4 9 5 3 10 8 11 6 2 7 6] |
| 12-137 | [0, 1, 5, 2, 8, 10, 3, 11, 9, 4, 7, 6] | [ 1 4 9 6 2 5 8 10 7 3 11 6] |
| 12-138 | [0, 1, 5, 2, 8, 7, 10, 3, 11, 9, 4, 6] | [ 1 4 9 6 11 3 5 8 10 7 2 6] |
| 12-139 | [0, 1, 5, 2, 9, 11, 4, 10, 8, 7, 3, 6] | [ 1 4 9 7 2 5 6 10 11 8 3 6] |
| 12-140 | [0, 1, 5, 2, 9, 11, 7, 10, 4, 3, 8, 6] | [ 1 4 9 7 2 8 3 6 11 5 10 6] |
| 12-141 | [0, 1, 5, 2, 9, 11, 10, 4, 7, 3, 8, 6] | [ 1 4 9 7 2 11 6 3 8 5 10 6] |
| 12-142L | [0, 1, 5, 2, 9, 7, 3, 8, 11, 10, 4, 6] | [ 1 4 9 7 10 8 5 3 11 2 6 6] |
| 12-143 | [0, 1, 5, 2, 10, 8, 3, 9, 11, 4, 7, 6] | [ 1 4 9 8 10 7 6 2 5 3 11 6] |
| 12-144 | [0, 1, 5, 3, 8, 11, 7, 2, 4, 10, 9, 6] | [ 1 4 10 5 3 8 7 2 6 11 9 6] |
| 12-145 | [0, 1, 5, 3, 8, 2, 4, 11, 7, 10, 9, 6] | [ 1 4 10 5 6 2 7 8 3 11 9 6] |
| 12-146 | [0, 1, 5, 3, 8, 2, 11, 4, 9, 7, 10, 6] | [ 1 4 10 5 6 9 8 5 3 11 6] |
| 12-147 | [0, 1, 5, 3, 9, 2, 4, 11, 8, 7, 10, 6] | [ 1 4 10 6 5 2 7 9 11 3 8 6] |
| 12-148 | [0, 1, 5, 3, 10, 9, 2, 8, 11, 7, 4, 6] | [ 1 4 10 7 11 5 6 3 8 9 2 6] |
| 12-149 | [0, 1, 5, 3, 11, 4, 7, 2, 8, 10, 9, 6] | [ 1 4 10 8 5 3 7 6 2 11 9 6] |
| 12-150 | [0, 1, 5, 3, 11, 8, 2, 7, 10, 9, 4, 6] | [ 1 4 10 8 9 6 5 3 11 2 6] |
| 12-151 | [0, 1, 5, 4, 7, 2, 8, 10, 3, 11, 9, 6] | [ 1 4 11 3 7 6 2 5 8 10 9 6] |
| 12-152 | [0, 1, 5, 4, 7, 2, 10, 3, 9, 11, 8, 6] | [ 1 4 11 3 7 8 5 6 2 9 10 6] |
| 12-153L | [0, 1, 5, 4, 7, 2, 11, 9, 3, 8, 10, 6] | [ 1 4 11 3 7 9 10 6 5 2 8 6] |
| 12-154L | [0, 1, 5, 4, 10, 8, 3, 11, 2, 7, 9, 6] | [ 1 4 11 6 10 7 8 3 5 2 9 6] |
| 12-155 | [0, 1, 5, 4, 11, 2, 10, 7, 9, 3, 8, 6] | [ 1 4 11 7 3 8 9 2 6 5 10 6] |
| 12-156 | [0, 1, 5, 4, 11, 9, 2, 8, 10, 7, 3, 6] | [ 1 4 11 7 10 5 6 2 9 8 3 6] |
| 12-157 | [0, 1, 5, 4, 2, 10, 7, 9, 3, 8, 11, 6] | [ 1 4 11 10 8 9 2 6 5 3 7 6] |
| 12-158 | [0, 1, 7, 9, 2, 10, 8, 5, 4, 11, 3, 6] | [ 1 6 2 5 8 10 9 11 7 4 3 6] |
| 12-159 | [0, 1, 7, 10, 2, 4, 9, 8, 5, 3, 11, 6] | [ 1 6 3 4 2 5 11 9 10 8 7 6] |

Table I. Listing of all the 918 AIS generators in prime form

| Label | Row | Intervals |
| --- | --- | --- |
| 12-160 | [0, 1, 7, 10, 2, 9, 8, 5, 3, 11, 4, 6] | [ 1 6 3 4 7 11 9 10 8 5 2 6] |
| 12-161 | [0, 1, 7, 10, 3, 5, 9, 8, 4, 2, 11, 6] | [ 1 6 3 5 2 4 11 8 10 9 7 6] |
| 12-162 | [0, 1, 7, 10, 5, 9, 2, 4, 3, 11, 8, 6] | [ 1 6 3 7 4 5 2 11 8 9 10 6] |
| 12-163 | [0, 1, 7, 10, 8, 4, 3, 5, 9, 2, 11, 6] | [ 1 6 3 10 8 11 2 4 5 9 7 6] |
| 12-164 | [0, 1, 7, 10, 8, 5, 9, 2, 4, 3, 11, 6] | [ 1 6 3 10 9 4 5 2 11 8 7 6] |
| 12-165 | [0, 1, 7, 10, 9, 2, 4, 8, 5, 3, 11, 6] | [ 1 6 3 11 5 2 4 9 10 8 7 6] |
| 12-166 | [0, 1, 7, 10, 9, 5, 2, 4, 11, 3, 8, 6] | [ 1 6 3 11 8 9 2 7 4 5 10 6] |
| 12-167 | [0, 1, 7, 11, 2, 4, 9, 8, 5, 3, 10, 6] | [ 1 6 4 3 2 5 11 9 10 7 8 6] |
| 12-168 | [0, 1, 7, 11, 2, 10, 8, 3, 5, 4, 9, 6] | [ 1 6 4 3 8 10 7 2 11 5 9 6] |
| 12-169 | [0, 1, 7, 11, 10, 3, 5, 8, 4, 2, 9, 6] | [ 1 6 4 11 5 2 3 8 10 7 9 6] |
| 12-170 | [0, 1, 7, 2, 5, 3, 11, 4, 8, 10, 9, 6] | [ 1 6 7 3 10 8 5 4 2 11 9 6] |
| 12-171 | [0, 1, 7, 2, 5, 11, 8, 10, 3, 11, 9, 6] | [ 1 6 7 3 11 4 2 5 8 10 9 6] |
| 12-172 | [0, 1, 7, 2, 10, 3, 5, 4, 8, 11, 9, 6] | [ 1 6 7 8 5 2 11 4 3 10 9 6] |
| 12-173 | [0, 1, 7, 2, 10, 8, 11, 3, 5, 4, 9, 6] | [ 1 6 7 8 10 3 4 2 11 5 9 6] |
| 12-174 | [0, 1, 7, 2, 10, 8, 11, 4, 3, 5, 9, 6] | [ 1 6 7 8 10 3 5 11 2 4 9 6] |
| 12-175 | [0, 1, 7, 2, 10, 8, 5, 9, 11, 4, 3, 6] | [ 1 6 7 8 10 9 4 2 5 11 3 6] |
| 12-176 | [0, 1, 7, 2, 10, 8, 5, 4, 9, 11, 3, 6] | [ 1 6 7 8 10 9 11 5 2 4 3 6] |
| 12-177 | [0, 1, 7, 2, 10, 9, 11, 4, 8, 5, 3, 6] | [ 1 6 7 8 11 2 5 4 9 10 3 6] |
| 12-178 | [0, 1, 7, 2, 11, 4, 8, 10, 9, 5, 3, 6] | [ 1 6 7 9 5 4 2 11 8 10 3 6] |
| 12-179 | [0, 1, 7, 2, 11, 5, 9, 4, 8, 10, 3, 6] | [ 1 6 7 9 10 8 5 4 2 11 3 6] |
| 12-180 | [0, 1, 7, 3, 8, 10, 2, 5, 4, 11, 9, 6] | [ 1 6 8 5 2 4 3 11 7 10 9 6] |
| 12-181 | [0, 1, 7, 3, 10, 2, 5, 4, 9, 11, 8, 6] | [ 1 6 8 7 4 3 11 5 2 9 10 6] |
| 12-182 | [0, 1, 7, 3, 10, 8, 5, 4, 9, 11, 2, 6] | [ 1 6 8 7 10 9 11 5 2 3 4 6] |
| 12-183 | [0, 1, 7, 4, 2, 11, 9, 5, 4, 8, 10, 6] | [ 1 6 9 4 2 7 11 8 5 4 3 10 8 7 6] |
| 12-184 | [0, 1, 7, 4, 9, 11, 3, 2, 10, 5, 8, 6] | [ 1 6 9 5 2 4 11 8 7 3 10 6] |
| 12-185 | [0, 1, 7, 4, 9, 8, 10, 2, 5, 3, 11, 6] | [ 1 6 9 5 11 2 4 3 10 8 7 6] |
| 12-186 | [0, 1, 7, 4, 9, 8, 10, 5, 3, 11, 2, 6] | [ 1 6 9 5 11 2 7 10 8 3 4 6] |
| 12-187 | [0, 1, 7, 4, 11, 2, 10, 3, 5, 9, 8, 6] | [ 1 6 9 7 3 8 5 2 4 11 10 6] |
| 12-188 | [0, 1, 7, 4, 11, 9, 5, 8, 10, 3, 2, 6] | [ 1 6 9 7 10 8 3 2 5 11 4 6] |
| 12-189 | [0, 1, 7, 4, 2, 5, 9, 8, 10, 3, 11, 6] | [ 1 6 9 10 3 4 11 2 5 8 7 6] |
| 12-190 | [0, 1, 7, 4, 2, 9, 8, 11, 3, 5, 10, 6] | [ 1 6 9 10 7 11 3 4 2 5 8 6] |
| 12-191 | [0, 1, 7, 4, 3, 10, 2, 5, 9, 8, 11, 6] | [ 1 6 9 10 8 5 2 4 11 3 7 6] |
| 12-192 | [0, 1, 7, 4, 3, 5, 9, 2, 10, 8, 11, 6] | [ 1 6 9 11 2 4 5 8 10 3 7 6] |
| 12-193 | [0, 1, 7, 5, 8, 3, 11, 10, 2, 4, 9, 6] | [ 1 6 10 3 7 8 11 4 2 5 9 6] |
| 12-194 | [0, 1, 7, 5, 10, 2, 9, 11, 8, 4, 3, 6] | [ 1 6 10 5 4 7 2 9 8 11 3 6] |
| 12-195 | [0, 1, 7, 5, 2, 4, 9, 11, 3, 8, 10, 6] | [ 1 6 10 9 2 5 11 3 4 7 8 6] |
| 12-196 | [0, 1, 7, 5, 2, 10, 9, 11, 4, 8, 3, 6] | [ 1 6 10 9 8 11 2 5 4 7 3 6] |
| 12-197 | [0, 1, 7, 5, 4, 8, 10, 3, 11, 2, 9, 6] | [ 1 6 10 11 4 2 5 8 3 7 9 6] |
| 12-198P | [0, 1, 8, 10, 2, 11, 5, 4, 9, 7, 3, 6] | [ 1 7 2 4 9 6 11 5 10 8 3 6] |
| 12-199 | [0, 1, 8, 10, 2, 11, 7, 5, 4, 9, 3, 6] | [ 1 7 2 4 9 8 10 11 5 6 3 6] |
| 12-200 | [0, 1, 8, 10, 4, 3, 11, 2, 7, 5, 9, 6] | [ 1 7 2 6 11 8 3 5 10 4 9 6] |
| 12-201 | [0, 1, 8, 10, 7, 5, 4, 9, 3, 11, 2, 6] | [ 1 7 2 9 10 11 5 6 8 3 4 6] |
| 12-202 | [0, 1, 8, 10, 9, 3, 11, 4, 7, 5, 2, 6] | [ 1 7 2 11 6 8 5 3 10 9 4 6] |
| 12-203 | [0, 1, 8, 11, 3, 9, 2, 10, 7, 5, 4, 6] | [ 1 7 3 4 6 5 8 9 10 11 2 6] |
| 12-204 | [0, 1, 8, 11, 4, 10, 7, 9, 5, 3, 2, 6] | [ 1 7 3 5 6 9 2 8 10 11 4 6] |
| 12-205 | [0, 1, 8, 11, 4, 10, 9, 7, 3, 5, 2, 6] | [ 1 7 3 5 6 11 10 8 2 9 4 6] |
| 12-206 | [0, 1, 8, 11, 5, 4, 2, 10, 3, 7, 9, 6] | [ 1 7 3 6 11 10 8 5 4 2 9 6] |
| 12-207L | [0, 1, 8, 11, 7, 5, 10, 2, 4, 3, 9, 6] | [ 1 7 3 8 10 5 4 2 11 6 9 6] |
| 12-208 | [0, 1, 8, 11, 9, 5, 7, 4, 10, 3, 2, 6] | [ 1 7 3 10 8 2 9 6 5 11 4 6] |
| 12-209 | [0, 1, 8, 11, 9, 5, 2, 4, 10, 3, 7, 6] | [ 1 7 3 10 8 9 2 6 5 4 11 6] |
| 12-210 | [0, 1, 8, 11, 10, 3, 2, 9, 5, 7, 4, 6] | [ 1 7 3 11 5 6 8 2 9 10 4 6] |
| 12-211 | [0, 1, 8, 11, 10, 4, 9, 7, 3, 5, 2, 6] | [ 1 7 3 11 6 5 10 8 2 9 4 6] |
| 12-212 | [0, 1, 8, 11, 10, 4, 2, 7, 3, 5, 9, 6] | [ 1 7 3 11 6 10 5 8 2 4 9 6] |
| 12-213 | [0, 1, 8, 2, 5, 3, 7, 4, 9, 11, 10, 6] | [ 1 7 6 3 10 4 9 5 2 11 8 6] |
| 12-214L | [0, 1, 8, 2, 10, 3, 5, 4, 7, 11, 9, 6] | [ 1 7 6 8 5 3 2 4 3 10 9 6] |
| 12-215 | [0, 1, 8, 2, 10, 7, 5, 9, 11, 4, 3, 6] | [ 1 7 6 8 9 10 4 2 5 11 3 6] |
| 12-216 | [0, 1, 8, 2, 10, 7, 5, 4, 9, 11, 3, 6] | [ 1 7 6 8 9 10 11 5 2 4 3 6] |
| 12-217 | [0, 1, 8, 2, 11, 3, 5, 4, 9, 7, 10, 6] | [ 1 7 6 9 4 2 11 5 10 3 8 6] |
| 12-218 | [0, 1, 8, 2, 11, 4, 3, 5, 9, 7, 10, 6] | [ 1 7 6 9 5 11 2 4 10 3 8 6] |
| 12-219 | [0, 1, 8, 4, 10, 3, 2, 5, 7, 11, 9, 6] | [ 1 7 8 6 5 11 3 2 4 10 9 6] |
| 12-220 | [0, 1, 8, 4, 2, 5, 7, 11, 10, 3, 9, 6] | [ 1 7 8 10 3 2 4 11 5 6 9 6] |
| 12-221P | [0, 1, 8, 4, 2, 5, 11, 10, 3, 7, 9, 6] | [ 1 7 8 10 3 6 11 5 4 2 9 6] |
| 12-222L | [0, 1, 8, 5, 7, 10, 4, 3, 11, 2, 9, 6] | [ 1 7 9 2 3 11 6 5 10 8 3 6] |
| 12-223 | [0, 1, 8, 5, 7, 10, 9, 3, 11, 4, 2, 6] | [ 1 7 9 2 3 11 6 8 5 10 4 6] |
| 12-224L | [0, 1, 8, 5, 7, 11, 4, 2, 10, 9, 3, 6] | [ 1 7 9 2 4 5 10 8 11 6 3 6] |
| 12-225L | [0, 1, 8, 5, 7, 11, 9, 2, 10, 4, 3, 6] | [ 1 7 9 2 4 10 5 8 6 11 3 6] |
| 12-226 | [0, 1, 8, 5, 8, 9, 11, 10, 4, 2, 7, 3, 6] | [ 1 7 9 4 2 11 6 10 5 8 3 6] |
| 12-227 | [0, 1, 8, 5, 11, 3, 2, 4, 9, 7, 10, 6] | [ 1 7 9 6 4 11 2 5 10 3 8 6] |
| 12-228 | [0, 1, 8, 5, 11, 9, 2, 4, 3, 7, 10, 6] | [ 1 7 9 6 10 5 2 11 4 3 8 6] |
| 12-229 | [0, 1, 8, 5, 11, 10, 2, 4, 9, 7, 3, 6] | [ 1 7 9 6 11 4 2 5 10 8 3 6] |
| 12-230L | [0, 1, 8, 5, 3, 9, 2, 4, 7, 11, 10, 6] | [ 1 7 9 10 5 2 11 3 4 10 8 6] |
| 12-231L | [0, 1, 8, 5, 3, 11, 10, 4, 7, 9, 2, 6] | [ 1 7 9 10 8 11 6 3 2 5 4 6] |
| 12-232 | [0, 1, 8, 7, 9, 5, 10, 4, 2, 11, 3, 6] | [ 1 7 11 2 8 5 6 10 9 4 3 6] |
| 12-233 | [0, 1, 8, 7, 11, 9, 5, 2, 4, 10, 3, 6] | [ 1 7 11 4 10 8 9 2 6 5 3 6] |
| 12-234 | [0, 1, 8, 7, 5, 2, 10, 4, 9, 11, 3, 6] | [ 1 7 11 10 9 8 6 2 5 8 3 6] |
| 12-235 | [0, 1, 9, 11, 3, 2, 7, 4, 10, 5, 8, 6] | [ 1 8 2 4 11 5 9 6 7 3 10 6] |
| 12-236 | [0, 1, 9, 11, 4, 10, 7, 5, 8, 3, 2, 6] | [ 1 8 2 5 6 9 10 3 7 11 4 6] |
| 12-237L | [0, 1, 9, 11, 4, 3, 7, 10, 5, 2, 8, 6] | [ 1 8 2 5 11 4 3 7 9 6 10 6] |
| 12-238L | [0, 1, 9, 11, 4, 3, 10, 7, 5, 8, 2, 6] | [ 1 8 2 5 11 7 9 10 3 6 4 6] |
| 12-239 | [0, 1, 9, 2, 4, 3, 7, 10, 5, 11, 8, 6] | [ 1 8 5 2 11 4 3 7 6 9 10 6] |

| Label | Row | Intervals |
| --- | --- | --- |
| 12-240 | [0, 1, 9, 2, 4, 3, 7, 10, 8, 5, 11, 6] | [ 1 8 5 2 11 4 3 10 9 6 7 6] |
| 12-241L | [0, 1, 9, 2, 4, 3, 10, 7, 11, 5, 8, 6] | [ 1 8 5 2 11 7 9 4 6 3 10 6] |
| 12-242 | [0, 1, 9, 2, 5, 11, 3, 10, 8, 7, 4, 6] | [ 1 8 5 3 6 4 7 10 11 9 2 6] |
| 12-243 | [0, 1, 9, 2, 5, 11, 10, 8, 3, 7, 4, 6] | [ 1 8 5 3 6 11 10 7 4 9 2 6] |
| 12-244L | [0, 1, 9, 2, 8, 10, 7, 5, 4, 11, 3, 6] | [ 1 8 5 6 2 9 10 11 7 4 3 6] |
| 12-245L | [0, 1, 9, 2, 8, 11, 3, 10, 7, 5, 4, 6] | [ 1 8 5 6 3 4 7 9 10 11 2 6] |
| 12-246L | [0, 1, 9, 2, 8, 11, 10, 5, 3, 7, 4, 6] | [ 1 8 5 6 3 11 7 10 4 9 2 6] |
| 12-247 | [0, 1, 9, 3, 5, 10, 8, 7, 4, 11, 2, 6] | [ 1 8 6 2 5 10 11 9 7 3 4 6] |
| 12-248 | [0, 1, 9, 3, 8, 10, 7, 5, 4, 11, 2, 6] | [ 1 8 6 5 2 9 10 11 7 3 4 6] |
| 12-249 | [0, 1, 9, 3, 8, 11, 10, 5, 7, 4, 2, 6] | [ 1 8 6 5 3 11 7 2 9 10 4 6] |
| 12-250L | [0, 1, 9, 4, 7, 11, 10, 3, 5, 2, 8, 6] | [ 1 8 7 3 4 11 5 2 9 6 10 6] |
| 12-251 | [0, 1, 9, 4, 7, 5, 11, 10, 8, 3, 7, 6] | [ 1 8 7 3 10 6 9 2 5 11 4 6] |
| 12-252 | [0, 1, 9, 4, 10, 3, 2, 5, 7, 11, 8, 6] | [ 1 8 7 6 5 11 3 2 4 9 10 6] |
| 12-253S | [0, 1, 9, 4, 2, 5, 11, 8, 10, 3, 7, 6] | [ 1 8 7 10 3 6 9 2 5 4 11 6] |
| 12-254 | [0, 1, 9, 4, 2, 8, 5, 7, 11, 10, 3, 6] | [ 1 8 7 10 6 9 2 4 11 5 3 6] |
| 12-255S | [0, 1, 9, 4, 2, 11, 5, 8, 10, 3, 7, 6] | [ 1 8 7 10 9 6 3 2 5 4 11 6] |
| 12-256L | [0, 1, 9, 4, 3, 5, 10, 7, 11, 2, 8, 6] | [ 1 8 7 11 2 5 9 4 3 6 10 6] |
| 12-257 | [0, 1, 9, 4, 3, 8, 10, 7, 5, 11, 2, 6] | [ 1 8 7 11 5 2 9 10 6 3 4 6] |
| 12-258 | [0, 1, 9, 7, 10, 2, 4, 3, 8, 5, 11, 6] | [ 1 8 10 3 4 2 11 5 9 6 7 6] |
| 12-259 | [0, 1, 9, 7, 10, 3, 5, 2, 11, 4, 8, 6] | [ 1 8 10 3 5 2 11 4 6 9 7 6] |
| 12-260L | [0, 1, 9, 7, 10, 3, 5, 4, 11, 8, 2, 6] | [ 1 8 10 3 5 2 11 7 9 6 4 6] |
| 12-261 | [0, 1, 9, 7, 10, 3, 2, 4, 8, 5, 11, 6] | [ 1 8 10 3 5 11 2 4 9 6 7 6] |
| 12-262L | [0, 1, 9, 7, 11, 8, 3, 2, 5, 10, 4, 6] | [ 1 8 10 4 9 7 11 3 5 6 2 6] |
| 12-263 | [0, 1, 9, 7, 2, 4, 8, 5, 11, 10, 3, 6] | [ 1 8 10 7 2 4 9 6 5 11 3 6] |
| 12-264P | [0, 1, 9, 7, 2, 11, 5, 4, 8, 10, 3, 6] | [ 1 8 10 7 9 6 11 4 2 5 3 6] |
| 12-265 | [0, 1, 9, 7, 4, 8, 10, 3, 2, 5, 11, 6] | [ 1 8 10 9 4 2 5 11 3 6 7 6] |
| 12-266 | [0, 1, 9, 7, 4, 11, 5, 8, 10, 3, 2, 6] | [ 1 8 10 9 7 6 3 2 5 11 4 6] |
| 12-267L | [0, 1, 9, 7, 4, 11, 5, 10, 8, 3, 2, 6] | [ 1 8 10 9 7 11 5 6 3 2 4 6] |
| 12-268 | [0, 1, 9, 7, 4, 3, 5, 10, 2, 8, 11, 6] | [ 1 8 10 9 11 2 5 4 6 3 7 6] |
| 12-269 | [0, 1, 9, 7, 4, 3, 8, 10, 2, 5, 11, 6] | [ 1 8 10 9 11 5 2 4 3 6 7 6] |
| 12-270 | [0, 1, 9, 7, 4, 3, 8, 10, 5, 11, 2, 6] | [ 1 8 10 9 11 5 2 7 6 3 4 6] |
| 12-271 | [0, 1, 9, 8, 10, 2, 7, 4, 11, 5, 3, 6] | [ 1 8 11 2 4 5 9 7 6 10 3 6] |
| 12-272 | [0, 1, 9, 8, 10, 3, 7, 4, 2, 5, 11, 6] | [ 1 8 11 2 5 4 9 10 3 6 7 6] |
| 12-273 | [0, 1, 9, 8, 11, 5, 3, 10, 2, 7, 4, 6] | [ 1 8 11 3 6 10 7 4 5 9 2 6] |
| 12-274L | [0, 1, 9, 8, 2, 5, 3, 10, 7, 11, 4, 6] | [ 1 8 11 6 3 10 7 9 4 5 2 6] |
| 12-275 | [0, 1, 10, 2, 5, 11, 7, 9, 4, 3, 8, 6] | [ 1 9 4 3 6 8 2 7 11 5 10 6] |
| 12-276 | [0, 1, 10, 2, 5, 4, 11, 7, 9, 3, 8, 6] | [ 1 9 4 3 11 7 8 2 6 5 10 6] |
| 12-277 | [0, 1, 10, 2, 7, 9, 4, 3, 11, 5, 8, 6] | [ 1 9 4 5 2 7 11 8 6 3 10 6] |
| 12-278 | [0, 1, 10, 2, 7, 9, 8, 4, 11, 5, 3, 6] | [ 1 9 4 5 2 11 10 8 3 6 2 6] |
| 12-279 | [0, 1, 10, 2, 7, 5, 8, 4, 3, 9, 11, 6] | [ 1 9 4 5 10 3 8 11 6 2 7 6] |
| 12-280S | [0, 1, 10, 2, 9, 11, 5, 3, 8, 4, 7, 6] | [ 1 9 4 7 2 6 10 5 8 3 11 6] |
| 12-281P | [0, 1, 10, 2, 9, 11, 5, 4, 7, 3, 8, 6] | [ 1 9 4 7 2 6 11 3 8 5 10 6] |
| 12-282L | [0, 1, 10, 2, 9, 3, 8, 11, 7, 5, 4, 6] | [ 1 9 4 7 6 5 3 8 10 11 2 6] |
| 12-283L | [0, 1, 10, 2, 9, 7, 3, 8, 11, 5, 4, 6] | [ 1 9 4 7 10 8 5 3 6 11 2 6] |
| 12-284 | [0, 1, 10, 3, 5, 11, 9, 8, 4, 7, 2, 6] | [ 1 9 5 2 6 10 11 8 3 7 4 6] |
| 12-285 | [0, 1, 10, 3, 7, 9, 5, 4, 11, 2, 8, 6] | [ 1 9 5 4 2 8 11 7 3 6 10 6] |
| 12-286 | [0, 1, 10, 3, 7, 9, 5, 4, 2, 8, 11, 6] | [ 1 9 5 4 2 8 11 10 6 3 7 6] |
| 12-287 | [0, 1, 10, 3, 7, 9, 8, 4, 2, 5, 11, 6] | [ 1 9 5 4 2 11 8 10 3 6 7 6] |
| 12-288 | [0, 1, 10, 3, 7, 2, 5, 11, 9, 8, 4, 6] | [ 1 9 5 4 7 3 6 10 11 8 2 6] |
| 12-289L | [0, 1, 10, 3, 7, 2, 8, 11, 9, 5, 4, 6] | [ 1 9 5 4 7 6 3 10 8 11 2 6] |
| 12-290 | [0, 1, 10, 3, 7, 5, 11, 2, 9, 8, 4, 6] | [ 1 9 5 4 10 6 3 7 11 2 6] |
| 12-291 | [0, 1, 10, 3, 9, 11, 7, 2, 5, 4, 8, 6] | [ 1 9 5 6 2 8 7 3 11 4 10 6] |
| 12-292 | [0, 1, 10, 3, 9, 8, 4, 2, 5, 7, 11, 6] | [ 1 9 5 6 11 8 10 3 2 4 7 6] |
| 12-293 | [0, 1, 10, 3, 11, 2, 8, 7, 5, 9, 4, 6] | [ 1 9 5 8 3 6 11 10 4 7 2 6] |
| 12-294 | [0, 1, 10, 3, 11, 6, 5, 3, 2, 7, 10, 6] | [ 1 9 5 8 6 3 11 2 7 10 4 6] |
| 12-295 | [0, 1, 10, 3, 11, 9, 4, 8, 2, 5, 7, 6] | [ 1 9 5 8 10 7 4 6 3 2 11 6] |
| 12-296 | [0, 1, 10, 3, 2, 5, 11, 7, 9, 4, 8, 6] | [ 1 9 5 11 3 6 8 2 7 4 10 6] |
| 12-297 | [0, 1, 10, 3, 2, 8, 11, 7, 5, 9, 4, 6] | [ 1 9 5 11 6 3 8 10 4 7 2 6] |
| 12-298 | [0, 1, 10, 3, 2, 8, 4, 5, 9, 7, 11, 6] | [ 1 9 5 11 6 8 3 10 4 7 2 6] |
| 12-299 | [0, 1, 10, 3, 2, 9, 7, 11, 5, 8, 4, 6] | [ 1 9 5 11 7 10 4 6 3 8 2 6] |
| 12-300 | [0, 1, 10, 4, 8, 3, 11, 9, 2, 5, 7, 6] | [ 1 9 6 4 7 8 10 5 3 2 11 6] |
| 12-301 | [0, 1, 10, 4, 8, 7, 9, 2, 5, 3, 11, 6] | [ 1 9 6 4 11 2 5 3 10 8 7 6] |
| 12-302 | [0, 1, 10, 4, 9, 11, 7, 5, 8, 3, 2, 6] | [ 1 9 6 5 2 8 10 3 7 11 4 6] |
| 12-303 | [0, 1, 10, 4, 11, 2, 7, 3, 5, 9, 8, 6] | [ 1 9 6 7 3 5 8 2 4 11 10 6] |
| 12-304 | [0, 1, 10, 4, 11, 3, 2, 7, 9, 5, 8, 6] | [ 1 9 6 7 4 11 5 2 8 3 10 6] |
| 12-305 | [0, 1, 10, 4, 2, 7, 3, 5, 9, 8, 11, 6] | [ 1 9 6 10 5 8 2 4 11 3 7 6] |
| 12-306 | [0, 1, 10, 4, 3, 5, 9, 11, 10, 8, 7, 3, 6] | [ 1 9 6 11 5 3 10 8 2 7 8 3 6] |
| 12-307 | [0, 1, 10, 5, 7, 11, 9, 2, 8, 4, 3, 6] | [ 1 9 7 2 4 10 5 6 8 11 3 6] |
| 12-308P | [0, 1, 10, 5, 7, 3, 9, 8, 11, 4, 2, 6] | [ 1 9 7 2 8 6 11 3 5 10 4 6] |
| 12-309 | [0, 1, 10, 5, 8, 7, 9, 3, 11, 4, 2, 6] | [ 1 9 7 3 11 2 6 8 5 10 4 6] |
| 12-310 | [0, 1, 10, 5, 8, 7, 3, 11, 4, 2, 6] | [ 1 9 7 3 11 8 6 2 5 10 4 6] |
| 12-311L | [0, 1, 10, 5, 9, 11, 4, 7, 3, 2, 8, 6] | [ 1 9 7 4 2 5 3 8 11 6 10 6] |
| 12-312L | [0, 1, 10, 5, 9, 11, 4, 2, 8, 7, 3, 6] | [ 1 9 7 4 2 5 10 6 11 8 3 6] |
| 12-313L | [0, 1, 10, 5, 9, 7, 3, 2, 8, 11, 4, 6] | [ 1 9 7 4 10 8 11 6 3 5 2 6] |
| 12-314 | [0, 1, 10, 5, 11, 4, 9, 8, 2, 7, 3, 6] | [ 1 9 7 6 5 7 11 6 2 5 8 3 6] |
| 12-315L | [0, 1, 10, 5, 3, 11, 4, 7, 9, 8, 2, 6] | [ 1 9 7 10 8 5 3 2 11 6 4 6] |
| 12-316 | [0, 1, 10, 8, 11, 4, 3, 9, 5, 7, 2, 6] | [ 1 9 10 3 5 11 6 8 2 7 4 6] |
| 12-317 | [0, 1, 10, 8, 2, 7, 9, 5, 4, 11, 3, 6] | [ 1 9 10 6 5 2 8 11 7 4 3 6] |
| 12-318 | [0, 1, 10, 8, 2, 9, 5, 7, 11, 4, 3, 6] | [ 1 9 10 6 7 8 2 4 5 11 3 6] |
| 12-319S | [0, 1, 10, 8, 3, 11, 5, 9, 2, 4, 7, 6] | [ 1 9 10 7 8 6 4 5 2 3 11 6] |

Table I. Listing of all the 918 AIS generators in prime form (continued).

| Label | Row | Intervals | Label | Row | Intervals |
|---|---|---|---|---|---|
| 12-320P | [0, 1, 10, 8, 3, 11, 5, 4, 7, 9, 2, 6] | [ 1 9 10 7 8 6 11 3 2 5 4 6] | 12-400S | [0, 2, 3, 11, 4, 7, 1, 10, 5, 9, 8, 6] | [ 2 1 8 5 3 6 9 7 4 11 10 6] |
| 12-321P | [0, 1, 10, 8, 4, 9, 3, 2, 5, 7, 11, 6] | [ 1 9 10 8 5 6 11 3 2 4 7 6] | 12-401 | [0, 2, 3, 11, 4, 10, 1, 8, 7, 5, 9, 6] | [ 2 1 8 5 6 3 7 11 10 4 9 6] |
| 12-322 | [0, 1, 10, 8, 4, 11, 3, 9, 2, 5, 7, 6] | [ 1 9 10 8 7 4 6 5 3 2 11 6] | 12-402S | [0, 2, 3, 11, 4, 1, 7, 10, 5, 9, 8, 6] | [ 2 1 8 5 9 6 3 7 4 11 10 6] |
| 12-323 | [0, 1, 10, 8, 4, 3, 7, 9, 2, 5, 11, 6] | [ 1 9 10 8 11 4 2 5 3 7 4 10, 6] | 12-403 | [0, 2, 3, 11, 8, 7, 5, 9, 4, 10, 1, 6] | [ 2 1 8 9 11 10 4 7 3 6 5 6] |
| 12-324 | [0, 1, 10, 8, 4, 3, 9, 2, 5, 7, 11, 6] | [ 1 9 10 8 11 6 5 3 2 4 7 6] | 12-404 | [0, 2, 3, 11, 10, 8, 5, 9, 4, 7, 1, 6] | [ 2 1 8 11 10 9 4 7 3 6 5 6] |
| 12-325 | [0, 1, 10, 8, 7, 2, 4, 9, 5, 11, 3, 6] | [ 1 9 10 11 7 2 5 8 6 4 3 6] | 12-405 | [0, 2, 3, 1, 7, 11, 10, 5, 8, 4, 9, 6] | [ 2 1 10 6 4 11 7 3 8 5 9 6] |
| 12-326 | [0, 1, 10, 8, 7, 2, 5, 9, 3, 11, 4, 6] | [ 1 9 10 11 7 3 4 6 8 5 2 6] | 12-406 | [0, 2, 3, 1, 8, 4, 7, 11, 5, 10, 9, 6] | [ 2 1 10 7 8 3 4 6 5 11 9 6] |
| 12-327 | [0, 1, 10, 9, 11, 5, 3, 8, 4, 7, 2, 6] | [ 1 9 11 2 6 10 5 8 3 7 4 6] | 12-407L | [0, 2, 3, 1, 9, 8, 5, 10, 4, 7, 11, 6] | [ 2 1 10 8 11 9 5 6 3 4 7 6] |
| 12-328 | [0, 1, 10, 9, 3, 5, 8, 4, 2, 7, 11, 6] | [ 1 9 11 6 2 3 8 10 5 4 7 6] | 12-408 | [0, 2, 3, 1, 10, 4, 9, 5, 8, 7, 11, 6] | [ 2 1 10 9 6 5 8 3 11 4 7 6] |
| 12-329 | [0, 1, 10, 9, 3, 8, 4, 2, 5, 7, 11, 6] | [ 1 9 11 6 5 8 10 3 2 4 7 6] | 12-409 | [0, 2, 5, 9, 10, 4, 11, 8, 7, 3, 1, 6] | [ 2 3 4 1 6 7 9 11 8 10 5 6] |
| 12-330 | [0, 1, 10, 9, 5, 7, 11, 4, 2, 8, 3, 6] | [ 1 9 11 8 2 4 5 10 6 7 3 6] | 12-410 | [0, 2, 5, 9, 10, 8, 1, 7, 4, 3, 11, 6] | [ 2 3 4 1 10 5 6 9 11 8 7 6] |
| 12-331 | [0, 1, 10, 9, 5, 3, 7, 2, 8, 11, 4, 6] | [ 1 9 11 8 10 4 7 6 3 5 2 6] | 12-411 | [0, 2, 5, 9, 7, 8, 1, 10, 4, 3, 11, 6] | [ 2 3 4 10 1 5 9 6 11 8 7 6] |
| 12-332 | [0, 1, 10, 9, 5, 3, 8, 2, 4, 7, 11, 6] | [ 1 9 11 8 10 5 6 2 3 4 7 6] | 12-412 | [0, 2, 5, 9, 7, 8, 3, 11, 10, 4, 1, 6] | [ 2 3 4 10 1 7 8 11 6 9 5 6] |
| 12-333 | [0, 1, 10, 9, 7, 2, 5, 11, 3, 8, 4, 6] | [ 1 9 11 10 7 3 6 4 5 8 2 6] | 12-413L | [0, 2, 5, 10, 11, 7, 4, 3, 9, 1, 8, 6] | [ 2 3 5 1 8 9 11 6 4 7 10 6] |
| 12-334S | [0, 1, 11, 2, 10, 3, 9, 4, 8, 5, 7, 6] | [ 1 10 3 8 5 6 7 4 9 2 11 6] | 12-414 | [0, 2, 5, 10, 7, 8, 4, 3, 9, 1, 11, 6] | [ 2 3 5 9 1 8 11 6 4 10 7 6] |
| 12-335 | [0, 1, 11, 2, 10, 5, 9, 3, 8, 7, 4, 6] | [ 1 10 3 8 7 5 6 4 11 9 2 6] | 12-415 | [0, 2, 5, 10, 8, 9, 1, 4, 3, 7, 11, 6] | [ 2 3 5 10 1 4 6 9 11 8 7 6] |
| 12-336 | [0, 1, 11, 3, 5, 8, 7, 2, 10, 4, 9, 6] | [ 1 10 4 2 3 11 7 8 6 5 9 6] | 12-416S | [0, 2, 5, 10, 9, 1, 7, 3, 4, 11, 8, 6] | [ 2 3 5 11 4 6 8 1 7 9 10 6] |
| 12-337 | [0, 1, 11, 3, 8, 2, 10, 5, 4, 7, 9, 6] | [ 1 10 4 5 6 8 7 11 3 2 9 6] | 12-417 | [0, 2, 5, 10, 9, 3, 7, 8, 4, 1, 11, 6] | [ 2 3 5 11 6 4 1 8 9 10 7 6] |
| 12-338 | [0, 1, 11, 3, 9, 2, 10, 5, 8, 7, 4, 6] | [ 1 10 4 6 5 8 7 3 11 9 2 6] | 12-418 | [0, 2, 5, 11, 3, 10, 9, 7, 8, 4, 1, 6] | [ 2 3 6 4 7 11 10 1 8 9 5 6] |
| 12-339 | [0, 1, 11, 3, 10, 9, 5, 8, 2, 7, 4, 6] | [ 1 10 4 7 11 8 3 6 5 9 2 6] | 12-419 | [0, 2, 5, 11, 3, 1, 8, 7, 4, 9, 10, 6] | [ 2 3 6 4 10 7 11 9 5 1 8 6] |
| 12-340P | [0, 1, 11, 4, 7, 3, 9, 8, 10, 5, 2, 6] | [ 1 10 5 3 8 6 11 2 7 9 4 6] | 12-420L | [0, 2, 5, 11, 7, 4, 3, 10, 8, 9, 1, 6] | [ 2 3 6 8 9 11 7 10 1 4 5 6] |
| 12-341 | [0, 1, 11, 4, 7, 3, 2, 8, 10, 5, 9, 6] | [ 1 10 5 3 8 11 6 2 7 4 9 6] | 12-421 | [0, 2, 5, 11, 9, 4, 3, 7, 8, 1, 10, 6] | [ 2 3 6 10 7 11 4 1 5 9 8 6] |
| 12-342 | [0, 1, 11, 4, 8, 2, 5, 7, 3, 10, 9, 6] | [ 1 10 5 4 6 3 2 8 7 11 9 6] | 12-422L | [0, 2, 5, 11, 10, 7, 3, 4, 9, 1, 8, 6] | [ 2 3 6 11 9 8 1 5 4 7 10 6] |
| 12-343 | [0, 1, 11, 4, 8, 5, 7, 2, 10, 9, 3, 6] | [ 1 10 5 4 9 2 7 11 4 8 3, 6] | 12-423 | [0, 2, 5, 1, 7, 11, 4, 3, 10, 8, 9, 6] | [ 2 3 8 6 4 5 11 7 10 1 9 6] |
| 12-344 | [0, 1, 11, 4, 10, 2, 9, 8, 5, 7, 3, 6] | [ 1 10 5 6 4 7 11 9 2 8 3 6] | 12-424 | [0, 2, 5, 1, 7, 11, 10, 8, 3, 4, 9, 6] | [ 2 3 8 6 4 11 10 7 1 5 9 6] |
| 12-345 | [0, 1, 11, 4, 10, 7, 9, 5, 8, 3, 2, 6] | [ 1 10 5 6 9 2 8 3 7 11 4 6] | 12-425 | [0, 2, 5, 1, 10, 4, 3, 8, 9, 7, 11, 6] | [ 2 3 8 9 6 11 5 1 10 4 7 6] |
| 12-346 | [0, 1, 11, 4, 3, 9, 5, 8, 10, 7, 2, 6] | [ 1 10 5 11 6 8 3 2 9 7 4 6] | 12-426L | [0, 2, 5, 3, 8, 9, 1, 7, 4, 11, 10, 6] | [ 2 3 10 5 1 4 6 9 7 11 8 6] |
| 12-347 | [0, 1, 11, 4, 8, 5, 7, 2, 9, 5, 8, 2, 6] | [ 1 10 5 11 7 9 2 5 11 6, 6] | 12-427 | [0, 2, 5, 3, 8, 9, 4, 1, 7, 11, 10, 6] | [ 2 3 10 5 1 7 9 6 4 11 8 6] |
| 12-348 | [0, 1, 11, 5, 8, 4, 3, 10, 2, 7, 9, 6] | [ 1 10 6 3 8 11 7 4 5 2 9 6] | 12-428 | [0, 2, 5, 3, 10, 11, 4, 8, 7, 1, 9, 6] | [ 2 3 10 7 1 5 4 11 6 8 9 6] |
| 12-349 | [0, 1, 11, 5, 8, 4, 3, 10, 7, 9, 2, 6] | [ 1 10 6 3 8 11 7 9 2 5 4 6] | 12-429 | [0, 2, 5, 4, 8, 9, 7, 1, 10, 3, 11, 6] | [ 2 3 11 4 1 10 6 9 5 8 7 6] |
| 12-350L | [0, 1, 11, 5, 8, 7, 3, 10, 2, 4, 9, 6] | [ 1 10 6 3 11 8 7 4 2 5 9 6] | 12-430 | [0, 2, 5, 4, 8, 1, 7, 3, 10, 11, 9, 6] | [ 2 3 11 4 5 6 8 7 1 10 9 6] |
| 12-351 | [0, 1, 11, 5, 8, 4, 3, 2, 4, 9, 8, 3, 7, 10, 6] | [ 1 10 6 9 2 5 11 7 3 4 10 6] | 12-431 | [0, 2, 5, 4, 8, 3, 11, 9, 10, 7, 1, 6] | [ 2 3 11 4 7 8 10 1 9 7 6] |
| 12-352 | [0, 1, 11, 5, 4, 7, 3, 8, 10, 2, 9, 6] | [ 1 10 6 11 3 8 5 2 4 7 9 6] | 12-432S | [0, 2, 5, 4, 9, 1, 7, 3, 10, 11, 8, 6] | [ 2 3 11 5 4 6 8 7 1 9 10 6] |
| 12-353 | [0, 1, 11, 7, 10, 2, 8, 3, 5, 4, 9, 6] | [ 1 10 8 3 4 6 7 2 11 5 9 6] | 12-433 | [0, 2, 5, 4, 10, 7, 8, 3, 11, 9, 1, 6] | [ 2 3 11 6 9 1 7 8 10 4 5 6] |
| 12-354 | [0, 1, 11, 7, 10, 2, 9, 3, 8, 5, 4, 6] | [ 1 10 8 3 4 7 6 5 9 11 2 6] | 12-434 | [0, 2, 5, 4, 10, 7, 3, 8, 9, 1, 11, 6] | [ 2 3 11 6 9 8 5 1 4 10 7 6] |
| 12-355L | [0, 1, 11, 7, 10, 3, 5, 4, 8, 2, 9, 6] | [ 1 10 8 3 5 2 11 4 6 7 9 6] | 12-435 | [0, 2, 5, 4, 11, 3, 8, 9, 7, 1, 10, 6] | [ 2 3 11 7 4 5 1 10 6 9 8 6] |
| 12-356 | [0, 1, 11, 7, 10, 3, 2, 8, 5, 9, 4, 6] | [ 1 10 8 3 5 11 6 9 4 7 2 6] | 12-436 | [0, 2, 5, 4, 11, 3, 9, 7, 8, 1, 10, 6] | [ 2 3 11 7 4 6 10 1 5 9 8 6] |
| 12-357 | [0, 1, 11, 7, 10, 4, 9, 8, 3, 5, 2, 6] | [ 1 10 8 3 6 5 11 7 2 9 4 6] | 12-437 | [0, 2, 5, 4, 11, 9, 3, 7, 8, 1, 10, 6] | [ 2 3 11 7 10 6 4 1 5 9 8 6] |
| 12-358 | [0, 1, 11, 7, 2, 4, 8, 5, 10, 9, 3, 6] | [ 1 10 8 7 2 4 9 5 11 6 3 6] | 12-438 | [0, 2, 7, 8, 11, 5, 9, 4, 3, 1, 10, 6] | [ 2 5 1 3 6 4 7 11 10 9 8 6] |
| 12-359 | [0, 1, 11, 7, 2, 5, 4, 8, 10, 3, 9, 6] | [ 1 10 8 7 3 11 4 2 5 6 9 6] | 12-439 | [0, 2, 7, 8, 4, 11, 10, 1, 5, 3, 9, 6] | [ 2 5 1 8 7 11 3 4 10 6 9 6] |
| 12-360 | [0, 1, 11, 8, 10, 3, 9, 5, 4, 7, 2, 6] | [ 1 10 9 2 5 6 8 11 3 7 4 6] | 12-440L | [0, 2, 7, 8, 4, 1, 11, 10, 5, 9, 3, 6] | [ 2 5 1 8 9 10 11 7 4 6 3 6] |
| 12-361 | [0, 1, 11, 8, 10, 5, 4, 7, 3, 9, 2, 6] | [ 1 10 9 2 7 11 3 8 6 5 4 6] | 12-441 | [0, 2, 7, 8, 4, 3, 10, 1, 5, 11, 9, 6] | [ 2 5 1 8 11 7 3 4 6 10 9 6] |
| 12-362L | [0, 1, 11, 8, 3, 7, 9, 2, 5, 4, 10, 6] | [ 1 10 9 7 4 2 5 3 11 8 6, 6] | 12-442 | [0, 2, 7, 8, 4, 3, 10, 1, 11, 5, 9, 6] | [ 2 5 1 8 11 7 3 10 6 9 6] |
| 12-363S | [0, 1, 11, 8, 4, 9, 3, 10, 2, 5, 7, 6] | [ 1 10 9 8 5 6 7 4 3 2 11 6] | 12-443 | [0, 2, 7, 8, 5, 9, 3, 1, 4, 11, 10, 6] | [ 2 5 1 9 4 6 10 3 7 11 8 6] |
| 12-364 | [0, 1, 11, 10, 7, 2, 5, 9, 3, 8, 4, 6] | [ 1 10 11 9 7 3 4 6 5 8 2 6] | 12-444 | [0, 2, 7, 8, 5, 11, 9, 1, 4, 3, 10, 6] | [ 2 5 1 9 6 10 4 3 11 7 8 6] |
| 12-365 | [0, 1, 11, 10, 7, 3, 8, 2, 5, 9, 4, 6] | [ 1 10 11 9 8 5 6 3 4 7 2 6] | 12-445 | [0, 2, 7, 8, 5, 3, 9, 1, 4, 11, 10, 6] | [ 2 5 1 9 10 6 4 3 7 11 8 6] |
| 12-366 | [0, 2, 3, 7, 10, 4, 9, 8, 5, 1, 11, 6] | [ 2 1 4 3 6 5 11 9 8 4 5 6] | 12-446S | [0, 2, 7, 10, 11, 5, 4, 8, 3, 1, 9, 6] | [ 2 5 3 1 4 6 8 11 9 7 10 6] |
| 12-367 | [0, 2, 3, 7, 10, 4, 11, 9, 8, 5, 1, 6] | [ 2 1 4 3 6 7 10 11 9 8 5 6] | 12-447 | [0, 2, 7, 10, 11, 5, 4, 8, 3, 1, 9, 6] | [ 2 5 3 1 6 11 4 7 10 8 9 6] |
| 12-368SL | [0, 2, 3, 7, 10, 5, 11, 4, 1, 9, 8, 6] | [ 2 1 4 3 7 6 5 9 8 11 10 6] | 12-448 | [0, 2, 7, 10, 4, 11, 3, 1, 9, 8, 5, 6] | [ 2 5 3 6 7 4 10 8 11 9 1 6] |
| 12-369P | [0, 2, 3, 7, 10, 5, 11, 9, 8, 4, 1, 6] | [ 2 1 4 3 7 6 10 11 8 9 5 6] | 12-449 | [0, 2, 7, 10, 4, 3, 11, 9, 1, 8, 5, 6] | [ 2 5 3 6 11 8 10 4 7 9 1 6] |
| 12-370L | [0, 2, 3, 7, 10, 9, 5, 11, 4, 1, 8, 6] | [ 2 1 4 3 11 8 6 5 9 7 10 6] | 12-450S | [0, 2, 7, 10, 9, 5, 11, 3, 4, 1, 8, 6] | [ 2 5 3 11 8 6 4 1 9 7 10 6] |
| 12-371 | [0, 2, 3, 7, 1, 4, 11, 9, 8, 5, 10, 6] | [ 2 1 4 6 3 7 10 11 9 5 8 6] | 12-451 | [0, 2, 7, 11, 5, 8, 9, 4, 3, 1, 10, 6] | [ 2 5 4 6 3 1 7 11 10 9 8 6] |
| 12-372 | [0, 2, 3, 7, 1, 11, 10, 5, 8, 4, 9, 6] | [ 2 1 4 6 10 11 7 3 8 5 9 6] | 12-452 | [0, 2, 7, 11, 5, 1, 4, 3, 10, 8, 9, 6] | [ 2 5 4 6 8 3 11 7 10 1 9 6] |
| 12-373L | [0, 2, 3, 7, 4, 9, 5, 11, 10, 1, 8, 6] | [ 2 1 4 9 5 8 6 11 3 7 10 6] | 12-453 | [0, 2, 7, 11, 5, 1, 10, 8, 9, 4, 3, 6] | [ 2 5 4 6 8 9 10 1 7 11 3 6] |
| 12-374 | [0, 2, 3, 7, 4, 11, 5, 1, 9, 3, 8, 5, 10, 7, 6] | [ 2 1 4 9 6 8 11 4 3 5 10 7 6] | 12-454L | [0, 2, 7, 11, 5, 4, 9, 1, 8, 10, 3, 6] | [ 2 5 4 6 11 9 8 11 7 10 5 6] |
| 12-375SL | [0, 2, 3, 7, 4, 11, 5, 10, 1, 9, 8, 6] | [ 2 1 4 9 7 6 5 3 8 11 10 6] | 12-455 | [0, 2, 7, 11, 8, 3, 1, 4, 10, 9, 5, 6] | [ 2 5 4 9 7 10 3 6 11 8 1 6] |
| 12-376 | [0, 2, 3, 7, 5, 10, 4, 1, 9, 8, 11, 6] | [ 2 1 4 10 5 6 9 8 11 3 7 6] | 12-456 | [0, 2, 7, 11, 10, 1, 8, 4, 5, 3, 9, 6] | [ 2 5 4 11 3 7 8 1 10 6 9 6] |
| 12-377 | [0, 2, 3, 7, 5, 11, 10, 1, 8, 4, 9, 6] | [ 2 1 4 10 6 11 3 7 8 5 9 6] | 12-457 | [0, 2, 7, 11, 10, 4, 1, 8, 9, 5, 3, 6] | [ 2 5 4 11 6 9 7 1 8 10 3 6] |
| 12-378 | [0, 2, 3, 7, 5, 1, 8, 11, 4, 10, 9, 6] | [ 2 1 4 10 8 7 3 5 6 11 9 6] | 12-458 | [0, 2, 7, 1, 5, 8, 4, 3, 10, 11, 9, 6] | [ 2 5 6 4 3 8 11 7 1 10 9 6] |
| 12-379 | [0, 2, 3, 7, 5, 1, 8, 11, 10, 4, 9, 6] | [ 2 1 4 10 8 7 3 11 6 5 9 6] | 12-459 | [0, 2, 7, 1, 9, 10, 8, 5, 4, 11, 3, 6] | [ 2 5 6 8 1 10 9 11 7 4 3 6] |
| 12-380 | [0, 2, 3, 7, 5, 1, 10, 4, 9, 8, 11, 6] | [ 2 1 4 10 8 9 6 5 11 3 7 6] | 12-460L | [0, 2, 7, 1, 10, 11, 9, 5, 4, 8, 3, 6] | [ 2 5 6 9 1 10 8 11 4 7 3 6] |
| 12-381P | [0, 2, 3, 8, 4, 7, 1, 11, 10, 5, 9, 6] | [ 2 1 5 8 3 6 10 11 7 4 9 6] | 12-461 | [0, 2, 7, 1, 10, 8, 9, 5, 4, 11, 3, 6] | [ 2 5 6 9 10 1 8 11 7 4 3 6] |
| 12-382 | [0, 2, 3, 8, 4, 11, 10, 1, 7, 5, 9, 6] | [ 2 1 5 8 7 11 3 6 10 4 5 6] | 12-462 | [0, 2, 7, 1, 10, 5, 9, 3, 4, 8, 11, 6] | [ 2 5 6 9 10 11 8 4 7 1 6] |
| 12-383 | [0, 2, 3, 8, 5, 9, 4, 7, 1, 11, 10, 6] | [ 2 1 5 9 4 7 3 6 10 11 8 6] | 12-463 | [0, 2, 7, 3, 9, 1, 4, 11, 10, 8, 5, 6] | [ 2 5 8 6 4 3 7 11 10 9 1 6] |
| 12-384 | [0, 2, 3, 8, 5, 9, 7, 1, 4, 11, 10, 6] | [ 2 1 5 9 4 10 6 3 7 11 8 6] | 12-464 | [0, 2, 7, 3, 10, 1, 11, 5, 4, 8, 9, 6] | [ 2 5 8 7 3 10 6 11 4 1 9 6] |
| 12-385 | [0, 2, 3, 8, 5, 1, 4, 10, 9, 7, 11, 6] | [ 2 1 5 9 8 3 6 11 10 4 7 6] | 12-465 | [0, 2, 7, 4, 5, 8, 3, 9, 1, 11, 10, 6] | [ 2 5 9 1 3 7 6 4 10 11 8 6] |
| 12-386 | [0, 2, 3, 8, 5, 4, 10, 1, 9, 7, 11, 6] | [ 2 1 5 9 11 6 3 8 10 4 7 6] | 12-466S | [0, 2, 7, 4, 5, 9, 3, 11, 10, 1, 8, 6] | [ 2 5 9 1 4 6 8 11 3 7 10 6] |
| 12-387 | [0, 2, 3, 8, 5, 4, 11, 9, 1, 7, 10, 6] | [ 2 1 5 9 11 7 10 4 6 3 8 6] | 12-467P | [0, 2, 7, 4, 5, 9, 3, 1, 8, 11, 10, 6] | [ 2 5 9 1 4 6 10 7 3 11 8 6] |
| 12-388 | [0, 2, 3, 9, 1, 4, 11, 8, 7, 5, 10, 6] | [ 2 1 6 4 3 7 9 11 10 5 8 6] | 12-468 | [0, 2, 7, 4, 5, 3, 9, 1, 8, 11, 10, 6] | [ 2 5 9 1 10 6 4 7 3 11 8 6] |
| 12-389 | [0, 2, 3, 9, 4, 7, 11, 10, 8, 5, 1, 6] | [ 2 1 6 7 3 4 11 10 9 8 5 6] | 12-469S | [0, 2, 7, 4, 3, 11, 5, 9, 10, 1, 8, 6] | [ 2 5 9 11 8 6 4 1 3 7 10 6] |
| 12-390 | [0, 2, 3, 9, 4, 8, 11, 10, 7, 5, 1, 6] | [ 2 1 6 7 4 3 11 9 10 8 5 6] | 12-470 | [0, 2, 7, 4, 3, 1, 8, 11, 5, 10, 9, 6] | [ 2 5 9 11 10 7 3 6 4 1 9 6] |
| 12-391 | [0, 2, 3, 9, 7, 11, 10, 5, 8, 4, 1, 6] | [ 2 1 6 10 4 11 7 3 8 9 5 6] | 12-471 | [0, 2, 7, 5, 8, 9, 1, 10, 4, 3, 11, 6] | [ 2 5 10 3 1 4 9 6 11 8 7 6] |
| 12-392 | [0, 2, 3, 9, 7, 4, 8, 11, 10, 5, 1, 6] | [ 2 1 6 10 9 4 3 11 7 8 5 6] | 12-472 | [0, 2, 7, 5, 8, 4, 3, 9, 1, 10, 11, 6] | [ 2 5 10 3 8 11 6 4 9 1 7 6] |
| 12-393S | [0, 2, 3, 10, 1, 5, 11, 7, 4, 9, 8, 6] | [ 2 1 7 3 4 6 8 9 5 11 10 6] | 12-473 | [0, 2, 7, 5, 11, 8, 9, 1, 4, 3, 10, 6] | [ 2 5 10 6 9 1 4 3 11 7 8 6] |
| 12-394 | [0, 2, 3, 10, 4, 7, 11, 9, 8, 5, 1, 6] | [ 2 1 7 6 3 4 10 11 9 8 5 6] | 12-474 | [0, 2, 7, 5, 1, 8, 11, 3, 4, 10, 9, 6] | [ 2 5 10 8 7 3 4 1 6 11 9 6] |
| 12-395S | [0, 2, 3, 10, 7, 11, 5, 1, 4, 9, 8, 6] | [ 2 1 7 9 4 6 8 3 5 11 10 6] | 12-475 | [0, 2, 8, 11, 3, 10, 9, 7, 4, 5, 1, 6] | [ 2 6 3 4 7 11 10 9 1 8 5 6] |
| 12-396 | [0, 2, 3, 10, 8, 11, 7, 1, 5, 4, 9, 6] | [ 2 1 7 10 3 8 6 4 11 5 9 6] | 12-476L | [0, 2, 8, 11, 7, 4, 5, 3, 10, 9, 1, 6] | [ 2 6 3 8 9 1 10 7 11 4 5 6] |
| 12-397 | [0, 2, 3, 10, 8, 7, 11, 5, 1, 4, 9, 6] | [ 2 1 7 10 11 4 6 8 3 5 9 6] | 12-477 | [0, 2, 8, 11, 10, 5, 9, 7, 3, 4, 1, 6] | [ 2 6 3 11 7 4 10 8 1 9 5 6] |
| 12-398 | [0, 2, 3, 10, 9, 7, 11, 5, 8, 4, 1, 6] | [ 2 1 7 11 10 4 6 3 8 9 5 6] | 12-478L | [0, 2, 8, 1, 4, 3, 10, 11, 7, 5, 9, 6] | [ 2 6 5 3 11 7 1 8 10 4 9 6] |
| 12-399 | [0, 2, 3, 10, 9, 7, 4, 8, 11, 5, 1, 6] | [ 2 1 7 11 10 9 4 3 6 8 5 6] | 12-479 | [0, 2, 8, 1, 4, 3, 10, 7, 11, 9, 5, 6] | [ 2 6 5 3 11 7 9 4 10 8 1 6] |

Table I. Listing of all the 918 AIS generators in prime form (continued).

| Label | Row | Intervals |
|---|---|---|
| 12-480L | [0, 2, 8, 1, 5, 3, 4, 11, 7, 10, 9, 6] | [ 2 6 5 4 10 1 7 8 3 11 9 6] |
| 12-481L | [0, 2, 8, 1, 5, 4, 7, 3, 10, 11, 9, 6] | [ 2 6 5 4 11 3 8 7 1 10 9 6] |
| 12-482L | [0, 2, 8, 1, 5, 4, 11, 9, 10, 7, 3, 6] | [ 2 6 5 4 11 7 10 1 9 8 3 6] |
| 12-483L | [0, 2, 8, 1, 9, 10, 7, 5, 4, 11, 3, 6] | [ 2 6 5 8 1 9 10 11 7 4 3 6] |
| 12-484 | [0, 2, 8, 1, 10, 11, 7, 5, 9, 4, 3, 6] | [ 2 6 5 9 1 8 10 4 7 11 3 6] |
| 12-485 | [0, 2, 8, 1, 10, 9, 5, 3, 4, 7, 11, 6] | [ 2 6 5 9 11 8 10 1 3 4 7 6] |
| 12-486L | [0, 2, 8, 3, 7, 10, 11, 9, 5, 4, 1, 6] | [ 2 6 7 4 3 1 10 8 11 9 5 6] |
| 12-487 | [0, 2, 8, 5, 9, 7, 3, 4, 11, 10, 1, 6] | [ 2 6 9 4 10 7 11 3 6 5 8 6] |
| 12-488L | [0, 2, 8, 5, 3, 4, 11, 7, 10, 9, 1, 6] | [ 2 6 9 10 1 7 8 3 11 4 5 6] |
| 12-489L | [0, 2, 8, 5, 4, 7, 3, 10, 11, 9, 1, 6] | [ 2 6 9 11 3 8 7 1 10 4 5 6] |
| 12-490 | [0, 2, 8, 7, 10, 3, 11, 9, 4, 1, 5, 6] | [ 2 6 11 3 5 8 10 7 9 4 1 6] |
| 12-491 | [0, 2, 8, 7, 10, 5, 1, 11, 3, 4, 9, 6] | [ 2 6 11 3 7 8 10 4 1 5 9 6] |
| 12-492L | [0, 2, 8, 7, 11, 4, 5, 3, 10, 1, 9, 6] | [ 2 6 11 4 5 1 10 7 3 8 9 6] |
| 12-493L | [0, 2, 8, 7, 11, 4, 1, 9, 10, 5, 3, 6] | [ 2 6 11 4 5 9 8 1 7 10 3 6] |
| 12-494L | [0, 2, 8, 7, 11, 9, 4, 5, 1, 10, 3, 6] | [ 2 6 11 4 10 7 1 8 9 5 3 6] |
| 12-495L | [0, 2, 8, 7, 3, 10, 1, 11, 4, 5, 9, 6] | [ 2 6 11 8 7 3 10 5 1 4 9 6] |
| 12-496 | [0, 2, 8, 7, 4, 9, 10, 1, 5, 3, 11, 6] | [ 2 6 11 9 5 1 3 4 10 8 7 6] |
| 12-497L | [0, 2, 8, 7, 4, 9, 10, 5, 1, 11, 3, 6] | [ 2 6 11 9 5 1 7 8 10 4 3 6] |
| 12-498S | [0, 2, 9, 10, 1, 5, 11, 7, 4, 3, 8, 6] | [ 2 7 1 3 4 6 8 9 11 5 10 6] |
| 12-499 | [0, 2, 9, 10, 7, 11, 5, 8, 4, 3, 1, 6] | [ 2 7 1 9 4 6 8 11 1 10 5 6] |
| 12-500S | [0, 2, 9, 10, 7, 11, 5, 1, 4, 3, 8, 6] | [ 2 7 1 9 4 6 8 3 11 5 10 6] |
| 12-501 | [0, 2, 9, 10, 8, 1, 5, 11, 7, 4, 3, 6] | [ 2 7 1 10 5 4 6 8 9 11 3 6] |
| 12-502 | [0, 2, 9, 10, 8, 5, 1, 7, 11, 4, 3, 6] | [ 2 7 1 10 9 8 6 4 5 11 3 6] |
| 12-503SL | [0, 2, 9, 1, 4, 5, 11, 10, 7, 3, 8, 6] | [ 2 7 4 3 1 6 11 9 8 5 10 6] |
| 12-504 | [0, 2, 9, 1, 4, 10, 3, 11, 8, 7, 5, 6] | [ 2 7 4 3 6 5 8 9 11 10 1 6] |
| 12-505L | [0, 2, 9, 1, 4, 3, 11, 5, 10, 7, 8, 6] | [ 2 7 4 3 11 8 6 5 9 1 10 6] |
| 12-506 | [0, 2, 9, 1, 7, 4, 5, 3, 8, 11, 10, 6] | [ 2 7 4 6 9 1 10 5 3 11 8 6] |
| 12-507 | [0, 2, 9, 1, 7, 5, 10, 11, 8, 4, 3, 6] | [ 2 7 4 6 10 6 5 1 9 8 7 6] |
| 12-508SL | [0, 2, 9, 1, 10, 11, 5, 4, 7, 3, 8, 6] | [ 2 7 4 9 1 6 11 3 8 5 10 6] |
| 12-509L | [0, 2, 9, 1, 10, 3, 11, 5, 4, 7, 8, 6] | [ 2 7 4 9 5 8 6 11 3 1 10 6] |
| 12-510 | [0, 2, 9, 1, 10, 4, 3, 8, 11, 7, 5, 6] | [ 2 7 4 9 6 11 5 3 8 10 1 6] |
| 12-511 | [0, 2, 9, 1, 11, 5, 10, 7, 8, 4, 3, 6] | [ 2 7 4 10 6 5 9 1 8 4 3 6] |
| 12-512 | [0, 2, 9, 1, 11, 7, 8, 5, 10, 4, 3, 6] | [ 2 7 4 10 8 1 9 5 6 11 3 6] |
| 12-513 | [0, 2, 9, 1, 11, 7, 8, 5, 4, 10, 3, 6] | [ 2 7 4 10 8 1 9 11 6 5 3 6] |
| 12-514 | [0, 2, 9, 1, 11, 7, 10, 4, 3, 8, 5, 6] | [ 2 7 4 10 8 3 6 11 5 9 1 6] |
| 12-515 | [0, 2, 9, 1, 11, 10, 4, 7, 3, 8, 5, 6] | [ 2 7 4 11 10 1 6 8 5 3 8 6] |
| 12-516 | [0, 2, 9, 3, 7, 4, 5, 8, 1, 11, 10, 6] | [ 2 7 6 4 9 1 3 5 10 11 8 6] |
| 12-517 | [0, 2, 9, 5, 3, 4, 8, 11, 10, 7, 1, 6] | [ 2 7 8 10 1 4 3 11 9 6 5 6] |
| 12-518 | [0, 2, 9, 5, 3, 7, 8, 11, 10, 4, 1, 6] | [ 2 7 8 10 4 1 3 11 6 9 5 6] |
| 12-519S | [0, 2, 9, 5, 3, 4, 7, 1, 10, 11, 3, 8, 6] | [ 2 7 8 11 3 6 9 1 4 5 10 6] |
| 12-520 | [0, 2, 9, 5, 4, 10, 7, 8, 11, 3, 1, 6] | [ 2 7 8 11 6 9 1 3 4 10 5 6] |
| 12-521 | [0, 2, 9, 5, 4, 10, 7, 8, 1, 11, 3, 6] | [ 2 7 8 11 6 9 1 5 10 4 3 6] |
| 12-522S | [0, 2, 9, 5, 4, 1, 7, 10, 11, 3, 8, 6] | [ 2 7 8 11 9 6 3 1 4 5 10 6] |
| 12-523 | [0, 2, 9, 7, 8, 4, 1, 5, 11, 10, 3, 6] | [ 2 7 10 1 8 9 4 6 11 5 3 6] |
| 12-524 | [0, 2, 9, 7, 10, 4, 3, 11, 8, 1, 5, 6] | [ 2 7 10 3 6 11 8 9 5 4 1 6] |
| 12-525 | [0, 2, 9, 7, 1, 5, 10, 11, 8, 4, 3, 6] | [ 2 7 10 6 4 5 1 9 8 11 3 6] |
| 12-526 | [0, 2, 9, 7, 3, 4, 8, 5, 11, 10, 1, 6] | [ 2 7 10 8 1 4 9 6 11 3 5 6] |
| 12-527L | [0, 2, 9, 7, 3, 8, 11, 10, 4, 1, 5, 6] | [ 2 7 10 8 5 3 11 6 9 4 1 6] |
| 12-528L | [0, 2, 9, 8, 11, 3, 4, 10, 7, 5, 1, 6] | [ 2 7 11 3 4 1 6 9 10 8 5 6] |
| 12-529 | [0, 2, 9, 8, 11, 3, 4, 1, 7, 5, 10, 6] | [ 2 7 11 3 4 1 9 6 10 5 8 6] |
| 12-530 | [0, 2, 9, 8, 11, 3, 1, 7, 4, 5, 10, 6] | [ 2 7 11 3 4 10 6 9 1 5 8 6] |
| 12-531 | [0, 2, 9, 8, 11, 4, 5, 3, 7, 1, 10, 6] | [ 2 7 11 3 5 1 10 4 6 9 8 6] |
| 12-532 | [0, 2, 9, 8, 11, 4, 10, 7, 3, 1, 5, 6] | [ 2 7 11 3 5 6 9 8 10 4 1 6] |
| 12-533 | [0, 2, 9, 8, 11, 7, 4, 10, 3, 1, 5, 6] | [ 2 7 11 3 8 9 6 5 10 4 1 6] |
| 12-534 | [0, 2, 9, 8, 4, 5, 10, 7, 1, 11, 3, 6] | [ 2 7 11 8 1 5 9 6 10 4 3 6] |
| 12-535P | [0, 2, 9, 8, 4, 5, 10, 7, 1, 11, 3, 6] | [ 2 7 11 8 9 6 10 5 1 4 3 6] |
| 12-536P | [0, 2, 10, 11, 4, 1, 7, 5, 9, 8, 3, 6] | [ 2 8 1 5 9 6 10 4 11 7 3 6] |
| 12-537 | [0, 2, 10, 11, 4, 1, 8, 7, 5, 9, 3, 6] | [ 2 8 1 5 9 7 11 10 4 6 3 6] |
| 12-538 | [0, 2, 10, 11, 8, 1, 7, 5, 9, 4, 3, 6] | [ 2 8 1 9 5 6 10 4 7 11 3 6] |
| 12-539 | [0, 2, 10, 1, 7, 11, 9, 4, 3, 8, 5, 6] | [ 2 8 3 6 4 10 7 11 5 9 1 6] |
| 12-540 | [0, 2, 10, 1, 7, 5, 4, 11, 3, 8, 9, 6] | [ 2 8 3 6 10 11 7 4 5 1 9 6] |
| 12-541 | [0, 2, 10, 1, 8, 7, 11, 4, 5, 3, 9, 6] | [ 2 8 3 7 11 4 5 1 10 6 9 6] |
| 12-542 | [0, 2, 10, 1, 8, 7, 11, 5, 3, 4, 9, 6] | [ 2 8 3 7 11 4 4 6 9 1 5 9 6] |
| 12-543 | [0, 2, 10, 1, 8, 7, 5, 11, 3, 4, 9, 6] | [ 2 8 3 7 11 10 6 4 1 5 9 6] |
| 12-544 | [0, 2, 10, 3, 4, 1, 7, 5, 9, 8, 11, 6] | [ 2 8 5 1 9 6 10 4 11 3 7 6] |
| 12-545 | [0, 2, 10, 3, 7, 1, 4, 11, 9, 8, 5, 6] | [ 2 8 5 4 6 3 7 10 11 9 1 6] |
| 12-546 | [0, 2, 10, 3, 9, 1, 4, 11, 8, 7, 5, 6] | [ 2 8 5 6 4 3 7 9 11 10 1 6] |
| 12-547 | [0, 2, 10, 3, 1, 7, 4, 5, 9, 8, 11, 6] | [ 2 8 5 10 6 9 1 4 11 3 7 6] |
| 12-548 | [0, 2, 10, 5, 8, 7, 1, 11, 3, 4, 9, 6] | [ 2 8 7 3 11 6 10 4 1 5 9 6] |
| 12-549 | [0, 2, 10, 5, 4, 7, 8, 1, 11, 3, 9, 6] | [ 2 8 7 11 3 1 5 10 4 6 9 6] |
| 12-550 | [0, 2, 10, 5, 4, 7, 1, 11, 8, 9, 3, 6] | [ 2 8 7 11 3 4 9 1 6 10 5 6] |
| 12-551P | [0, 2, 10, 5, 4, 7, 1, 11, 3, 8, 9, 6] | [ 2 8 7 11 3 6 10 4 5 1 9 6] |
| 12-552 | [0, 2, 10, 7, 8, 11, 5, 9, 4, 3, 1, 6] | [ 2 8 9 1 3 6 4 7 11 10 5 6] |
| 12-553 | [0, 2, 10, 7, 8, 1, 5, 11, 9, 4, 3, 6] | [ 2 8 9 1 5 4 6 10 7 11 3 6] |
| 12-554 | [0, 2, 10, 7, 8, 1, 11, 4, 9, 3, 6] | [ 2 8 9 1 5 4 11 7 10 6 3 6] |
| 12-555 | [0, 2, 10, 7, 8, 1, 11, 5, 9, 4, 3, 6] | [ 2 8 9 1 5 10 6 4 7 11 3 6] |
| 12-556 | [0, 2, 10, 7, 11, 5, 8, 9, 4, 3, 1, 6] | [ 2 8 9 4 6 3 1 7 11 10 5 6] |
| 12-557 | [0, 2, 10, 7, 1, 5, 3, 4, 9, 8, 11, 6] | [ 2 8 9 6 4 10 1 5 11 3 7 6] |
| 12-558 | [0, 2, 10, 9, 7, 1, 11, 4, 5, 9, 8, 11, 6] | [ 2 8 9 6 10 1 5 4 11 7 3 6] |
| 12-559 | [0, 2, 10, 9, 1, 7, 4, 5, 3, 8, 11, 6] | [ 2 8 11 4 6 9 1 10 5 3 7 6] |
| 12-560 | [0, 2, 10, 9, 3, 7, 4, 5, 8, 1, 11, 6] | [ 2 8 11 6 4 9 1 3 5 10 7 6] |
| 12-561 | [0, 2, 10, 9, 4, 7, 8, 5, 11, 3, 1, 6] | [ 2 8 11 7 3 1 9 6 4 10 5 6] |
| 12-562 | [0, 2, 10, 9, 4, 7, 1, 11, 3, 8, 5, 6] | [ 2 8 11 7 3 6 10 4 5 9 1 6] |
| 12-563 | [0, 2, 10, 9, 7, 11, 5, 8, 3, 4, 1, 6] | [ 2 8 11 10 4 6 3 7 1 9 5 6] |
| 12-564 | [0, 2, 10, 9, 7, 1, 4, 11, 3, 8, 5, 6] | [ 2 8 11 10 6 3 7 4 5 9 1 6] |
| 12-565 | [0, 2, 11, 3, 10, 8, 9, 5, 4, 7, 1, 6] | [ 2 9 4 7 10 1 8 11 3 6 5 6] |
| 12-566 | [0, 2, 11, 3, 10, 8, 7, 1, 4, 9, 5, 6] | [ 2 9 4 7 10 11 6 3 5 8 1 6] |
| 12-567 | [0, 2, 11, 3, 1, 8, 7, 10, 4, 9, 5, 6] | [ 2 9 4 10 7 11 3 6 5 8 1 6] |
| 12-568 | [0, 2, 11, 4, 5, 9, 3, 1, 8, 7, 10, 6] | [ 2 9 5 1 4 6 10 7 11 3 8 6] |
| 12-569 | [0, 2, 11, 4, 5, 9, 8, 3, 1, 7, 10, 6] | [ 2 9 5 1 4 11 7 10 6 3 8 6] |
| 12-570 | [0, 2, 11, 4, 5, 3, 9, 1, 8, 7, 10, 6] | [ 2 9 5 1 10 6 4 7 11 3 8 6] |
| 12-571 | [0, 2, 11, 4, 8, 3, 1, 5, 7, 10, 9, 6] | [ 2 9 5 4 7 10 6 3 1 8 1 6] |
| 12-572 | [0, 2, 11, 4, 8, 7, 1, 9, 10, 5, 3, 6] | [ 2 9 5 4 11 6 8 1 7 10 3 6] |
| 12-573 | [0, 2, 11, 4, 10, 1, 9, 8, 3, 7, 5, 6] | [ 2 9 5 6 3 8 11 7 4 10 1 6] |
| 12-574S | [0, 2, 11, 4, 3, 7, 1, 9, 10, 5, 8, 6] | [ 2 9 5 11 4 6 8 1 7 3 10 6] |
| 12-575 | [0, 2, 11, 5, 7, 8, 1, 4, 3, 10, 9, 6] | [ 2 9 6 4 10 1 5 3 11 7 8 6] |
| 12-576 | [0, 2, 11, 5, 9, 7, 8, 4, 3, 10, 1, 6] | [ 2 9 6 4 10 1 8 11 3 7 5 6] |
| 12-577L | [0, 2, 11, 5, 10, 1, 9, 4, 3, 7, 8, 6] | [ 2 9 6 5 3 8 7 11 4 1 10 6] |
| 12-578 | [0, 2, 11, 5, 3, 4, 9, 1, 8, 7, 10, 6] | [ 2 9 6 10 1 5 4 7 11 3 8 6] |
| 12-579L | [0, 2, 11, 5, 3, 7, 10, 9, 8, 1, 6] | [ 2 9 6 10 1 8 7 3 10 4 5 6] |
| 12-580 | [0, 2, 11, 5, 4, 8, 9, 3, 10, 1, 6] | [ 2 9 6 11 4 1 10 8 7 3 5 6] |
| 12-581 | [0, 2, 11, 7, 10, 4, 9, 8, 3, 1, 5, 6] | [ 2 9 8 3 6 5 11 7 10 4 1 6] |
| 12-582 | [0, 2, 11, 7, 10, 5, 4, 8, 9, 3, 1, 6] | [ 2 9 8 3 7 11 4 1 6 10 5 6] |
| 12-583 | [0, 2, 11, 7, 1, 5, 10, 1, 7 3, 8, 6] | [ 2 9 8 6 4 5 10 1 7 11 3 6] |
| 12-584 | [0, 2, 11, 7, 1, 5, 4, 9, 10, 8, 3, 6] | [ 2 9 8 6 4 11 5 1 10 7 3 6] |
| 12-585 | [0, 2, 11, 7, 5, 8, 3, 4, 10, 9, 1, 6] | [ 2 9 8 10 3 7 1 6 11 4 5 6] |
| 12-586 | [0, 2, 11, 7, 5, 9, 8, 3, 4, 10, 1, 6] | [ 2 9 8 10 4 11 7 1 6 3 5 6] |
| 12-587 | [0, 2, 11, 7, 5, 4, 8, 9, 3, 10, 1, 6] | [ 2 9 8 10 11 4 7 1 6 3 5 6] |
| 12-588 | [0, 2, 11, 9, 10, 5, 8, 4, 3, 7, 1, 6] | [ 2 9 10 1 7 3 8 11 4 6 5 6] |
| 12-589 | [0, 2, 11, 9, 10, 5, 4, 8, 1, 7, 3, 6] | [ 2 9 10 1 7 11 4 5 6 8 3 6] |
| 12-590 | [0, 2, 11, 9, 8, 3, 4, 7, 1, 5, 10, 6] | [ 2 9 10 11 7 1 3 6 4 5 8 6] |
| 12-591L | [0, 2, 11, 9, 8, 3, 7, 1, 4, 5, 10, 6] | [ 2 9 10 11 7 4 6 3 1 5 8 6] |
| 12-592 | [0, 2, 11, 10, 1, 8, 4, 9, 3, 7, 5, 6] | [ 2 9 11 3 7 8 5 6 4 10 1 6] |
| 12-593S | [0, 2, 11, 10, 3, 7, 1, 9, 4, 5, 8, 6] | [ 2 9 11 5 4 6 8 7 1 3 10 6] |
| 12-594 | [0, 2, 11, 10, 3, 9, 1, 8, 4, 7, 5, 6] | [ 2 9 11 5 6 4 7 8 3 10 1 6] |
| 12-595L | [0, 2, 11, 10, 5, 1, 4, 9, 3, 7, 8, 6] | [ 2 9 11 7 8 3 5 6 4 3 10 1 6 ] (?) |
| 12-596 | [0, 2, 11, 10, 8, 3, 7, 1, 4, 9, 5, 6] | [ 2 9 11 10 7 4 6 3 5 8 1 6] |
| 12-597S | [0, 2, 1, 4, 9, 5, 11, 3, 10, 7, 8, 6] | [ 2 11 3 5 8 6 4 7 9 1 10 6] |
| 12-598 | [0, 2, 1, 4, 9, 7, 8, 5, 11, 3, 10, 6] | [ 2 11 3 5 10 1 9 6 4 7 8 6] |
| 12-599S | [0, 2, 1, 4, 11, 3, 9, 5, 10, 7, 8, 6] | [ 2 11 3 7 4 6 8 5 9 1 10 6] |
| 12-600P | [0, 2, 1, 4, 11, 3, 9, 7, 8, 5, 10, 6] | [ 2 11 3 7 4 6 10 1 9 5 8 6] |
| 12-601 | [0, 2, 1, 4, 11, 8, 9, 3, 7, 5, 10, 6] | [ 2 11 3 7 9 1 6 4 10 5 8 6] |
| 12-602 | [0, 2, 1, 4, 11, 3, 9, 7, 8, 5, 10, 6] | [ 2 11 3 7 10 6 4 1 9 5 8 6] |
| 12-603 | [0, 2, 1, 5, 8, 9, 7, 4, 10, 3, 11, 6] | [ 2 11 4 3 1 10 9 6 5 8 7 6] |
| 12-604 | [0, 2, 1, 5, 10, 4, 7, 8, 3, 11, 9, 6] | [ 2 11 4 5 6 3 1 7 8 10 9 6] |
| 12-605 | [0, 2, 1, 5, 10, 4, 11, 8, 9, 7, 3, 6] | [ 2 11 4 5 6 7 9 1 10 8 3 6] |
| 12-606 | [0, 2, 1, 5, 10, 7, 8, 4, 11, 9, 3, 6] | [ 2 11 4 5 9 1 8 7 10 6 3 6] |
| 12-607L | [0, 2, 1, 5, 11, 4, 7, 3, 10, 8, 9, 6] | [ 2 11 4 6 5 3 8 7 10 1 9 6] |
| 12-608 | [0, 2, 1, 5, 11, 7, 10, 8, 3, 4, 9, 6] | [ 2 11 4 6 8 3 10 7 1 5 9 6] |
| 12-609 | [0, 2, 1, 5, 11, 7, 4, 9, 10, 8, 3, 6] | [ 2 11 4 6 8 9 5 1 10 7 3 6] |
| 12-610 | [0, 2, 1, 5, 11, 7, 1, 5, 10, 8, 3, 6] | [ 2 11 4 6 9 7 1 5 10 8 3 6] |
| 12-611 | [0, 2, 1, 7, 10, 3, 11, 9, 4, 8, 5, 6] | [ 2 11 6 3 5 8 10 7 4 9 1 6] |
| 12-612L | [0, 2, 1, 7, 10, 5, 3, 11, 4, 8, 9, 6] | [ 2 11 6 3 7 10 8 5 4 1 9 6] |
| 12-613 | [0, 2, 1, 7, 10, 8, 3, 11, 4, 5, 9, 6] | [ 2 11 6 3 10 7 8 5 1 4 9 6] |
| 12-614 | [0, 2, 1, 7, 1, 4, 9, 8, 5, 11, 3, 6] | [ 2 11 6 4 9 8 5 1 7 10 3 6] |
| 12-615 | [0, 2, 1, 7, 3, 10, 8, 11, 4, 5, 9, 6] | [ 2 11 6 8 7 10 3 5 1 4 9 6] |
| 12-616 | [0, 2, 1, 8, 11, 3, 9, 7, 4, 5, 10, 6] | [ 2 11 7 3 4 6 10 9 1 5 8 6] |
| 12-617 | [0, 2, 1, 8, 11, 5, 3, 7, 4, 9, 10, 6] | [ 2 11 7 3 6 10 4 9 5 1 8 6] |
| 12-618 | [0, 2, 1, 8, 11, 7, 3, 10, 6, 4, 9, 5, 6] | [ 2 11 7 3 10 6 4 9 1 5 8 6] |
| 12-619 | [0, 2, 1, 8, 4, 5, 10, 7, 11, 9, 3, 6] | [ 2 11 7 8 1 5 9 4 10 6 3 6] |
| 12-620L | [0, 2, 1, 8, 4, 7, 5, 10, 11, 3, 9, 6] | [ 2 11 7 8 3 10 5 1 4 6 9 6] |
| 12-621 | [0, 2, 1, 8, 4, 9, 10, 7, 11, 5, 3, 6] | [ 2 11 7 8 5 1 9 4 6 10 3 6] |
| 12-622 | [0, 2, 1, 8, 4, 9, 10, 7, 1, 5, 3, 6] | [ 2 11 7 8 5 1 9 10 6 4 3 6] |
| 12-623 | [0, 2, 1, 8, 5, 11, 3, 4, 9, 7, 10, 6] | [ 2 11 7 9 6 4 1 5 10 3 8 6] |
| 12-624 | [0, 2, 1, 9, 10, 7, 5, 11, 4, 8, 3, 6] | [ 2 11 8 1 9 10 6 5 4 7 3 6] |
| 12-625 | [0, 2, 1, 9, 3, 7, 4, 5, 10, 8, 11, 6] | [ 2 11 8 6 4 9 1 5 10 3 7 6] |
| 12-626 | [0, 2, 1, 9, 7, 8, 5, 10, 4, 11, 3, 6] | [ 2 11 8 10 1 9 5 6 7 4 3 6] (?) |
| 12-627 | [0, 2, 1, 9, 7, 10, 3, 4, 8, 5, 11, 6] | [ 2 11 8 10 3 5 1 4 9 6 7 6] |
| 12-628 | [0, 2, 1, 9, 7, 10, 4, 11, 3, 8, 5, 6] | [ 2 11 8 10 3 6 7 4 5 9 1 6] |
| 12-629 | [0, 2, 1, 10, 3, 9, 4, 8, 11, 7, 5, 6] | [ 2 11 9 5 6 7 4 3 8 10 1 6] |
| 12-630S | [0, 2, 1, 10, 3, 11, 5, 9, 4, 7, 8, 6] | [ 2 11 9 5 8 6 4 7 3 1 10 6] |
| 12-631 | [0, 2, 1, 10, 4, 9, 5, 3, 7, 8, 11, 6] | [ 2 11 9 6 5 8 10 4 1 3 7 6] |
| 12-632S | [0, 2, 1, 10, 5, 9, 3, 11, 4, 7, 8, 6] | [ 2 11 9 7 4 6 8 5 3 1 10 6] |
| 12-633 | [0, 2, 1, 10, 5, 11, 4, 8, 9, 7, 3, 6] | [ 2 11 9 7 6 5 4 1 10 8 3 6] |
| 12-634 | [0, 2, 1, 11, 3, 9, 4, 7, 8, 5, 10, 6] | [ 2 11 10 4 6 7 3 1 9 5 8 6] |
| 12-635 | [0, 2, 1, 11, 4, 7, 8, 5, 9, 3, 10, 6] | [ 2 11 10 5 3 1 9 4 6 7 8 6] |
| 12-636 | [0, 2, 1, 11, 5, 8, 3, 7, 4, 9, 10, 6] | [ 2 11 10 6 3 7 4 9 5 1 8 6] |
| 12-637 | [0, 2, 1, 11, 7, 8, 5, 9, 4, 10, 3, 6] | [ 2 11 10 8 1 9 4 7 6 5 3 6] |
| 12-638 | [0, 2, 1, 11, 7, 10, 3, 9, 4, 8, 5, 6] | [ 2 11 10 8 3 5 6 7 4 9 1 6] |
| 12-639 | [0, 2, 1, 11, 8, 3, 7, 10, 4, 9, 5, 6] | [ 2 11 10 9 7 4 3 6 5 8 1 6] |

Table I. Listing of all the 918 AIS generators in prime form (continued).
7

| Label | Row | Intervals | Label | Row | Intervals |
|---|---|---|---|---|---|
| 12-640 | [0, 2, 1, 11, 8, 4, 9, 3, 7, 10, 5, 6] | [ 2 11 10 9 8 5 6 4 3 7 1 6] | 12-720 | [0, 3, 7, 4, 5, 11, 9, 2, 1, 8, 10, 6] | [ 3 4 9 1 6 10 5 11 7 2 8 6] |
| 12-641L | [0, 3, 4, 8, 10, 9, 5, 2, 7, 1, 11, 6] | [ 3 1 4 2 11 8 9 5 6 10 7 6] | 12-721 | [0, 3, 7, 4, 2, 8, 1, 9, 11, 10, 5, 6] | [ 3 4 9 10 6 5 8 2 11 7 1 6] |
| 12-642L | [0, 3, 4, 8, 10, 9, 7, 2, 11, 5, 1, 6] | [ 3 1 4 2 11 10 7 9 6 8 5 6] | 12-722 | [0, 3, 7, 4, 2, 9, 8, 10, 11, 5, 1, 6] | [ 3 4 9 10 7 11 2 1 6 8 5 6] |
| 12-643 | [0, 3, 4, 8, 1, 7, 5, 2, 9, 11, 10, 6] | [ 3 1 4 5 6 10 9 7 2 11 8 6] | 12-723 | [0, 3, 7, 4, 2, 1, 8, 9, 11, 5, 10, 6] | [ 3 4 9 10 11 7 1 2 6 5 8 6] |
| 12-644 | [0, 3, 4, 8, 1, 7, 5, 2, 10, 9, 11, 6] | [ 3 1 4 5 6 10 9 8 11 2 7 6] | 12-724 | [0, 3, 7, 5, 10, 11, 8, 2, 1, 9, 4, 6] | [ 3 4 10 5 1 9 6 11 8 7 2 6] |
| 12-645 | [0, 3, 4, 8, 2, 7, 5, 1, 10, 9, 11, 6] | [ 3 1 4 6 5 10 8 9 11 2 7 6] | 12-725 | [0, 3, 7, 5, 11, 4, 1, 2, 9, 8, 10, 6] | [ 3 4 10 6 5 9 1 7 11 2 8 6] |
| 12-646 | [0, 3, 4, 8, 2, 9, 11, 10, 7, 5, 1, 6] | [ 3 1 4 6 7 2 11 9 10 8 5 6] | 12-726 | [0, 3, 7, 5, 11, 8, 1, 2, 10, 9, 4, 6] | [ 3 4 10 6 9 5 1 8 11 7 2 6] |
| 12-647L | [0, 3, 4, 9, 11, 5, 1, 10, 8, 7, 2, 6] | [ 3 1 5 2 6 8 9 10 11 7 4 6] | 12-727 | [0, 3, 7, 5, 2, 1, 8, 9, 11, 4, 10, 6] | [ 3 4 10 9 11 7 1 2 5 6 8 6] |
| 12-648 | [0, 3, 4, 9, 11, 10, 7, 5, 1, 8, 2, 6] | [ 3 1 5 2 11 9 10 8 7 4 6] | 12-728 | [0, 3, 9, 1, 2, 4, 11, 8, 7, 5, 10, 6] | [ 3 6 4 1 2 7 9 11 10 5 8 6] |
| 12-649 | [0, 3, 4, 9, 11, 10, 8, 5, 1, 7, 2, 6] | [ 3 1 5 2 11 10 9 8 6 7 4 6] | 12-729 | [0, 3, 9, 1, 2, 7, 5, 4, 11, 8, 10, 6] | [ 3 6 4 1 5 10 11 7 9 2 8 6] |
| 12-650P | [0, 3, 4, 9, 7, 11, 5, 2, 1, 8, 10, 6] | [ 3 1 5 10 4 6 9 11 7 2 8 6] | 12-730 | [0, 3, 9, 1, 8, 7, 5, 2, 10, 11, 4, 6] | [ 3 6 4 7 11 10 9 8 1 5 2 6] |
| 12-651 | [0, 3, 4, 9, 7, 2, 1, 5, 11, 8, 10, 6] | [ 3 1 5 10 7 11 4 6 9 2 8 6] | 12-731 | [0, 3, 9, 1, 11, 8, 7, 2, 1, 5, 10, 6] | [ 3 6 4 10 9 11 7 2 1 5 8 6] |
| 12-652 | [0, 3, 4, 10, 2, 7, 9, 8, 5, 1, 11, 6] | [ 3 1 6 4 5 2 11 9 8 10 7 6] | 12-732 | [0, 3, 9, 1, 11, 10, 5, 2, 7, 8, 4, 6] | [ 3 6 4 10 11 7 9 5 1 8 2 6] |
| 12-653 | [0, 3, 4, 10, 2, 9, 11, 8, 7, 5, 1, 6] | [ 3 1 6 4 7 2 9 11 10 8 5 6] | 12-733 | [0, 3, 9, 4, 5, 2, 10, 8, 7, 11, 1, 6] | [ 3 6 7 1 9 8 10 11 4 2 5 6] |
| 12-654 | [0, 3, 4, 10, 5, 1, 11, 8, 7, 9, 2, 6] | [ 3 1 6 7 8 10 9 11 2 5 4 6] | 12-734L | [0, 3, 9, 4, 8, 10, 11, 7, 5, 2, 1, 6] | [ 3 6 7 4 2 1 8 10 9 11 5 6] |
| 12-655 | [0, 3, 4, 10, 8, 5, 9, 11, 7, 2, 1, 6] | [ 3 1 6 10 9 4 2 8 7 11 5 6] | 12-735 | [0, 3, 9, 4, 8, 7, 5, 2, 10, 11, 1, 6] | [ 3 6 7 4 11 10 9 8 1 2 5 6] |
| 12-656 | [0, 3, 4, 11, 5, 1, 10, 8, 7, 9, 2, 6] | [ 3 1 7 6 8 9 10 11 2 5 4 6] | 12-736 | [0, 3, 10, 11, 9, 5, 7, 4, 8, 2, 1, 6] | [ 3 7 1 10 8 2 9 4 6 11 5 6] |
| 12-657 | [0, 3, 4, 11, 5, 2, 10, 8, 7, 9, 1, 6] | [ 3 1 7 6 9 8 10 11 2 4 5 6] | 12-737 | [0, 3, 10, 11, 9, 5, 2, 4, 8, 7, 1, 6] | [ 3 7 1 10 8 9 2 4 11 6 5 6] |
| 12-658 | [0, 3, 4, 11, 10, 8, 5, 7, 1, 9, 2, 6] | [ 3 1 7 11 10 9 2 6 8 5 4 6] | 12-738 | [0, 3, 10, 2, 7, 1, 11, 8, 9, 5, 4, 6] | [ 3 7 4 5 6 10 9 1 8 11 2 6] |
| 12-659 | [0, 3, 4, 1, 8, 2, 7, 5, 9, 11, 10, 6] | [ 3 1 9 7 6 5 10 4 2 11 8 6] | 12-739 | [0, 3, 10, 2, 11, 9, 8, 1, 7, 5, 4, 6] | [ 3 7 4 9 10 11 8 1 2 7 5 6] |
| 12-660L | [0, 3, 4, 1, 9, 11, 5, 10, 8, 7, 2, 6] | [ 3 1 9 8 2 6 5 10 11 7 4 6] | 12-740 | [0, 3, 10, 2, 1, 9, 7, 8, 5, 11, 4, 6] | [ 3 7 4 11 8 10 1 9 6 5 2 6] |
| 12-661 | [0, 3, 4, 1, 9, 8, 10, 2, 7, 5, 11, 6] | [ 3 1 9 8 11 2 4 5 10 6 7 6] | 12-741L | [0, 3, 10, 4, 8, 9, 11, 7, 5, 2, 1, 6] | [ 3 7 6 4 1 2 8 10 9 11 5 6] |
| 12-662 | [0, 3, 4, 2, 7, 9, 1, 8, 5, 11, 10, 6] | [ 3 1 10 5 2 4 7 9 6 11 8 6] | 12-742 | [0, 3, 10, 4, 9, 11, 8, 7, 5, 1, 2, 6] | [ 3 7 6 5 2 9 11 10 8 1 4 6] |
| 12-663S | [0, 3, 4, 2, 7, 11, 5, 1, 8, 10, 9, 6] | [ 3 1 10 5 4 6 8 7 2 11 9 6] | 12-743 | [0, 3, 10, 4, 2, 7, 11, 1, 5, 8, 9, 6] | [ 3 7 6 10 5 4 2 8 11 9 1 6] |
| 12-664 | [0, 3, 4, 2, 7, 1, 8, 5, 9, 11, 10, 6] | [ 3 1 10 5 6 7 9 4 2 11 8 6] | 12-744 | [0, 3, 10, 4, 2, 11, 7, 9, 8, 1, 5, 6] | [ 3 7 6 10 9 8 2 11 5 4 1 6] |
| 12-665 | [0, 3, 4, 2, 8, 1, 10, 9, 5, 7, 11, 6] | [ 3 1 10 6 5 9 11 8 2 4 7 6] | 12-745 | [0, 3, 10, 7, 8, 1, 9, 11, 5, 4, 2, 6] | [ 3 7 9 1 5 8 2 6 11 10 4 6] |
| 12-666 | [0, 3, 4, 2, 9, 11, 8, 7, 1, 5, 10, 6] | [ 3 1 10 7 2 9 11 6 4 5 8 6] | 12-746 | [0, 3, 10, 8, 9, 2, 11, 7, 1, 5, 4, 6] | [ 3 7 10 1 5 9 8 6 4 11 2 6] |
| 12-667 | [0, 3, 4, 2, 9, 8, 10, 5, 11, 7, 1, 6] | [ 3 1 10 7 11 2 9 4 6 8 5 6] | 12-747 | [0, 3, 10, 8, 9, 2, 5, 10, 11, 7, 1, 6] | [ 3 7 10 1 5 11 4 6 8 9 2 6] |
| 12-668 | [0, 3, 4, 2, 10, 5, 9, 11, 8, 7, 1, 6] | [ 3 1 10 8 7 4 2 9 11 6 5 6] | 12-748 | [0, 3, 10, 8, 9, 5, 2, 1, 7, 11, 4, 6] | [ 3 7 10 1 8 9 11 6 4 5 2 6] |
| 12-669P | [0, 3, 4, 2, 10, 5, 11, 8, 7, 9, 1, 6] | [ 3 1 10 8 7 6 9 11 2 4 5 6] | 12-749 | [0, 3, 10, 8, 7, 9, 5, 11, 4, 1, 2, 6] | [ 3 7 10 11 2 8 6 5 9 1 4 6] |
| 12-670 | [0, 3, 4, 2, 10, 7, 9, 8, 1, 5, 11, 6] | [ 3 1 10 8 9 2 11 5 4 6 7 6] | 12-750S | [0, 3, 10, 8, 7, 11, 5, 1, 2, 4, 9, 6] | [ 3 7 10 11 4 6 8 1 2 5 9 6] |
| 12-671 | [0, 3, 5, 9, 4, 2, 10, 11, 8, 7, 1, 6] | [ 3 2 4 7 10 2 4 11 10 5 6] | 12-751 | [0, 3, 10, 8, 7, 11, 2, 1, 5, 9, 4, 6] | [ 3 7 10 11 6 8 2 5 9 1 4 6] |
| 12-672P | [0, 3, 5, 9, 8, 1, 7, 4, 2, 10, 11, 6] | [ 3 2 4 11 5 6 9 10 8 1 7 6] | 12-752 | [0, 3, 10, 8, 7, 4, 9, 11, 5, 1, 2, 6] | [ 3 7 10 11 9 5 2 6 8 4 1 6] |
| 12-673 | [0, 3, 5, 9, 8, 4, 2, 7, 1, 10, 11, 6] | [ 3 2 4 11 8 10 5 6 9 1 7 6] | 12-753 | [0, 3, 10, 9, 11, 7, 4, 2, 8, 1, 5, 6] | [ 3 7 11 2 8 9 10 6 5 4 1 6] |
| 12-674 | [0, 3, 5, 10, 11, 8, 7, 1, 9, 4, 2, 6] | [ 3 2 5 1 9 11 6 8 7 10 4 6] | 12-754 | [0, 3, 10, 9, 11, 8, 4, 2, 7, 1, 5, 6] | [ 3 7 11 2 9 8 10 5 6 4 1 6] |
| 12-675 | [0, 3, 5, 10, 11, 2, 9, 8, 4, 1, 7, 2, 6] | [ 3 2 5 1 10 11 9 8 4 1 7 2 6] | 12-755 | [0, 3, 10, 9, 1, 2, 7, 5, 11, 8, 4, 6] | [ 3 7 11 4 1 5 10 6 9 8 2 6] |
| 12-676L | [0, 3, 5, 10, 9, 4, 1, 11, 7, 8, 2, 6] | [ 3 2 5 11 7 9 10 8 1 6 4 6] | 12-756P | [0, 3, 10, 9, 1, 11, 5, 2, 7, 8, 4, 6] | [ 3 7 11 4 10 6 9 5 1 8 2 6] |
| 12-677 | [0, 3, 5, 11, 4, 8, 9, 7, 2, 1, 10, 6] | [ 3 2 6 5 4 1 10 7 11 9 8 6] | 12-757 | [0, 4, 5, 7, 10, 3, 1, 9, 8, 2, 11, 6] | [ 4 1 2 3 5 10 8 11 6 9 7 6] |
| 12-678L | [0, 3, 5, 11, 4, 1, 9, 10, 8, 7, 2, 6] | [ 3 2 6 5 9 8 1 10 11 7 4 6] | 12-758 | [0, 4, 5, 7, 10, 8, 1, 9, 3, 2, 11, 6] | [ 4 1 2 3 10 5 8 6 11 9 7 6] |
| 12-679 | [0, 3, 5, 11, 7, 8, 1, 10, 9, 4, 2, 6] | [ 3 2 6 8 1 5 9 11 7 10 4 6] | 12-759 | [0, 4, 5, 7, 1, 9, 2, 11, 10, 8, 3, 6] | [ 4 1 2 6 8 5 9 11 10 7 3 6] |
| 12-680 | [0, 3, 5, 11, 10, 2, 9, 7, 8, 4, 1, 6] | [ 3 2 6 11 4 7 10 1 8 9 5 6] | 12-760 | [0, 4, 5, 7, 3, 9, 2, 1, 10, 8, 11, 6] | [ 4 1 2 8 6 5 11 9 10 3 7 6] |
| 12-681 | [0, 3, 5, 11, 10, 7, 8, 1, 9, 4, 2, 6] | [ 3 2 6 11 9 1 5 8 7 10 4 6] | 12-761 | [0, 4, 5, 8, 10, 7, 3, 9, 2, 1, 11, 6] | [ 4 1 3 2 9 8 6 5 11 10 7 6] |
| 12-682 | [0, 3, 5, 11, 10, 7, 8, 4, 2, 9, 1, 6] | [ 3 2 6 11 9 1 8 10 7 4 5 6] | 12-762S | [0, 4, 5, 8, 1, 3, 9, 7, 2, 11, 10, 6] | [ 4 1 3 5 2 6 10 7 9 11 8 6] |
| 12-683 | [0, 3, 5, 1, 7, 4, 11, 10, 8, 9, 2, 6] | [ 3 2 8 6 9 7 11 10 1 5 4 6] | 12-763 | [0, 4, 5, 8, 2, 7, 3, 1, 10, 9, 11, 6] | [ 4 1 3 6 5 8 10 9 11 2 7 6] |
| 12-684 | [0, 3, 5, 1, 7, 4, 8, 4, 10, 9, 1, 11, 6] | [ 3 2 9 5 1 8 6 11 4 10 7 6] | 12-764 | [0, 4, 5, 8, 2, 9, 11, 10, 7, 3, 1, 6] | [ 4 1 3 6 7 2 11 9 8 10 5 6] |
| 12-685 | [0, 3, 5, 2, 10, 11, 9, 4, 8, 7, 1, 6] | [ 3 2 9 8 1 10 7 4 11 6 5 6] | 12-765 | [0, 4, 5, 11, 7, 9, 2, 1, 10, 8, 3, 6] | [ 4 1 6 8 2 5 11 9 10 7 3 6] |
| 12-686L | [0, 3, 5, 2, 1, 8, 3, 9, 7, 11, 4, 10, 6] | [ 3 2 9 11 7 10 4 5 6 6] | 12-766 | [0, 4, 5, 11, 8, 1, 3, 9, 7, 11, 4, 10, 6] | [ 4 1 6 9 5 2 11 7 10 8 3 6] |
| 12-687 | [0, 3, 5, 4, 8, 9, 2, 10, 7, 1, 11, 6] | [ 3 2 11 4 1 5 8 9 6 10 7 6] | 12-767 | [0, 4, 5, 11, 9, 2, 1, 8, 10, 7, 3, 6] | [ 4 1 6 10 5 11 7 2 9 8 3 6] |
| 12-688S | [0, 3, 5, 4, 8, 1, 7, 2, 2, 10, 11, 9, 6] | [ 3 2 11 4 5 6 7 8 1 10 9 6] | 12-768 | [0, 4, 5, 1, 7, 2, 11, 9, 8, 10, 3, 6] | [ 4 1 8 6 7 9 10 11 2 5 3 6] |
| 12-689 | [0, 3, 5, 4, 8, 1, 7, 2, 11, 9, 10, 6] | [ 3 2 11 4 5 6 7 9 10 1 8 6] | 12-769 | [0, 4, 5, 1, 11, 10, 7, 9, 2, 8, 3, 6] | [ 4 1 8 10 11 9 2 5 6 7 3 6] |
| 12-690P | [0, 3, 5, 5, 4, 11, 1, 7, 1, 10, 8, 9, 2, 6] | [ 3 2 11 7 8 6 9 10 1 5 4 6] | 12-770S | [0, 4, 5, 2, 7, 9, 3, 1, 8, 11, 10, 6] | [ 4 1 9 5 2 6 10 7 3 11 8 6] |
| 12-691 | [0, 3, 5, 4, 1, 8, 2, 7, 11, 9, 10, 6] | [ 3 2 11 9 7 6 5 4 10 1 8 6] | 12-771 | [0, 4, 5, 2, 7, 1, 9, 11, 10, 8, 3, 6] | [ 4 1 9 5 8 6 2 11 10 7 3 6] |
| 12-692 | [0, 3, 5, 4, 1, 9, 10, 8, 2, 7, 11, 6] | [ 3 2 11 9 8 1 10 6 5 4 7 6] | 12-772 | [0, 4, 5, 2, 7, 3, 9, 8, 10, 1, 11, 6] | [ 4 1 9 5 8 6 11 2 3 10 7 6] |
| 12-693 | [0, 3, 5, 4, 2, 8, 1, 9, 10, 7, 11, 6] | [ 3 2 11 10 6 5 8 1 9 4 7 6] | 12-773 | [0, 4, 5, 2, 8, 7, 10, 3, 1, 9, 11, 6] | [ 4 1 9 6 11 3 5 10 8 2 7 6] |
| 12-694 | [0, 3, 7, 8, 10, 4, 9, 5, 2, 1, 11, 6] | [ 3 4 1 2 6 5 7 9 11 10 7 6] | 12-774 | [0, 4, 5, 2, 7, 3, 9, 8, 1, 7, 3, 6] | [ 4 1 9 7 2 11 10 5 6 8 3 6] |
| 12-695 | [0, 3, 7, 8, 10, 4, 11, 9, 5, 2, 1, 6] | [ 3 4 1 2 6 7 10 8 9 11 5 6] | 12-775 | [0, 4, 5, 3, 9, 2, 11, 1, 8, 7, 10, 6] | [ 4 1 10 6 5 9 2 7 11 3 8 6] |
| 12-696S | [0, 3, 7, 8, 10, 5, 11, 4, 2, 1, 9, 6] | [ 3 4 1 2 7 6 5 10 11 8 9 6] | 12-776 | [0, 4, 5, 3, 11, 2, 9, 8, 10, 7, 1, 6] | [ 4 1 10 8 3 7 11 2 9 6 5 6] |
| 12-697 | [0, 3, 7, 8, 10, 9, 4, 2, 11, 5, 1, 6] | [ 3 4 1 2 11 7 10 9 6 8 5 6] | 12-777P | [0, 4, 7, 8, 10, 3, 9, 5, 2, 1, 11, 6] | [ 4 3 1 2 5 6 8 9 11 10 7 6] |
| 12-698P | [0, 3, 7, 8, 1, 11, 5, 2, 10, 9, 4, 6] | [ 3 4 1 5 10 6 9 8 11 2 7 6] | 12-778S | [0, 4, 7, 8, 3, 5, 11, 1, 2, 10, 9, 6] | [ 4 3 1 7 2 6 10 5 11 9 8 6] |
| 12-699 | [0, 3, 7, 8, 5, 10, 4, 2, 1, 9, 11, 6] | [ 3 4 1 9 5 6 10 11 8 2 7 6] | 12-779 | [0, 4, 7, 8, 5, 11, 9, 2, 1, 3, 10, 6] | [ 4 3 1 9 6 10 5 11 2 7 8 6] |
| 12-700P | [0, 3, 7, 9, 10, 5, 11, 8, 4, 2, 1, 6] | [ 3 4 2 1 7 6 9 8 10 1 5 6] | 12-780 | [0, 4, 7, 9, 10, 3, 2, 8, 5, 1, 11, 6] | [ 4 3 2 1 5 11 6 9 8 10 7 6] |
| 12-701 | [0, 3, 7, 9, 10, 8, 4, 11, 5, 2, 1, 6] | [ 3 4 2 1 10 8 7 6 9 11 5 6] | 12-781 | [0, 4, 7, 9, 10, 5, 3, 11, 8, 2, 1, 6] | [ 4 3 2 1 7 10 8 9 6 11 5 6] |
| 12-702 | [0, 3, 7, 9, 2, 8, 4, 1, 11, 10, 5, 6] | [ 3 4 2 5 6 8 9 10 11 7 1 6] | 12-782 | [0, 4, 7, 9, 2, 3, 8, 1, 11, 10, 5, 6] | [ 4 3 2 1 10 7 8 6 9 11 5 6] |
| 12-703 | [0, 3, 7, 9, 2, 1, 10, 8, 4, 5, 11, 6] | [ 3 4 2 5 11 9 10 8 1 6 7 6] | 12-783 | [0, 4, 7, 9, 10, 8, 3, 2, 11, 5, 1, 6] | [ 4 3 2 1 10 11 7 11 9 6 8 5 6] |
| 12-704 | [0, 3, 7, 9, 2, 1, 11, 8, 4, 10, 5, 6] | [ 3 4 2 5 11 10 9 8 6 7 1 6] | 12-784L | [0, 4, 7, 9, 5, 3, 8, 2, 1, 10, 11, 6] | [ 4 3 2 8 10 5 6 11 9 1 7 6] |
| 12-705 | [0, 3, 7, 9, 4, 5, 2, 8, 1, 11, 10, 6] | [ 3 4 2 7 1 9 6 5 10 11 8 6] | 12-785 | [0, 4, 7, 9, 5, 3, 10, 11, 8, 2, 1, 6] | [ 4 3 2 8 10 7 1 9 6 11 5 6] |
| 12-706 | [0, 3, 7, 9, 5, 4, 10, 8, 1, 2, 11, 6] | [ 3 4 2 8 11 6 10 5 1 9 7 6] | 12-786L | [0, 4, 7, 9, 5, 3, 2, 8, 11, 1, 10, 6] | [ 4 3 2 8 11 6 10 3 2 9 6 5 6] |
| 12-707 | [0, 3, 7, 9, 5, 4, 2, 8, 1, 10, 11, 6] | [ 3 4 2 8 11 10 6 5 9 1 7 6] | 12-787 | [0, 4, 7, 9, 8, 1, 2, 11, 5, 3, 10, 6] | [ 4 3 2 11 5 1 9 6 10 7 8 6] |
| 12-708 | [0, 3, 7, 1, 9, 2, 4, 11, 10, 8, 5, 6] | [ 3 4 6 8 5 2 7 11 10 9 1 6] | 12-788 | [0, 4, 7, 9, 8, 1, 11, 5, 2, 3, 10, 6] | [ 4 3 2 11 5 10 6 9 1 7 8 6] |
| 12-709 | [0, 3, 7, 1, 11, 4, 5, 2, 9, 8, 10, 6] | [ 3 4 6 10 5 1 9 7 11 2 8 6] | 12-789 | [0, 4, 7, 9, 8, 3, 1, 2, 11, 5, 10, 6] | [ 4 3 2 11 7 10 1 9 6 5 8 6] |
| 12-710 | [0, 3, 7, 1, 11, 8, 9, 2, 10, 5, 4, 6] | [ 3 4 6 10 9 1 5 8 7 11 2 6] | 12-790 | [0, 4, 7, 1, 8, 10, 3, 2, 11, 5, 9, 6] | [ 4 3 6 7 2 5 11 9 6 10 7 8 6] |
| 12-711 | [0, 3, 7, 2, 4, 5, 11, 9, 8, 1, 10, 6] | [ 3 4 7 2 1 6 10 11 5 9 8 6] | 12-791 | [0, 4, 7, 1, 9, 2, 11, 10, 8, 3, 5, 6] | [ 4 3 6 8 5 9 11 10 7 2 1 6] |
| 12-712S | [0, 3, 7, 2, 4, 5, 11, 10, 8, 1, 9, 6] | [ 3 4 7 2 1 6 11 10 5 8 9 6] | 12-792 | [0, 4, 7, 1, 11, 8, 10, 3, 2, 9, 5, 6] | [ 4 3 6 10 9 2 5 11 7 8 1 6] |
| 12-713L | [0, 3, 7, 2, 8, 1, 10, 11, 9, 5, 4, 6] | [ 3 4 7 6 5 9 1 10 8 11 2 6] | 12-793S | [0, 4, 7, 2, 3, 5, 11, 9, 8, 1, 10, 6] | [ 4 3 7 1 2 6 10 11 5 9 8 6] |
| 12-714 | [0, 3, 7, 2, 1, 10, 8, 9, 11, 5, 4, 6] | [ 3 4 7 8 10 1 2 6 11 9 5 6] | 12-794 | [0, 4, 7, 2, 1, 10, 8, 9, 11, 5, 4, 6] | [ 4 3 7 9 11 10 5 2 6 8 1 6] |
| 12-715 | [0, 3, 7, 2, 1, 9, 10, 8, 5, 11, 4, 6] | [ 3 4 7 11 8 1 10 9 6 5 4 6] | 12-795 | [0, 4, 7, 2, 1, 3, 8, 5, 11, 9, 10, 6] | [ 4 3 11 2 5 9 6 10 1 8 6] |
| 12-716 | [0, 3, 7, 2, 1, 9, 11, 4, 10, 8, 5, 6] | [ 3 4 7 11 8 2 5 6 10 9 1 6] | 12-796 | [0, 4, 7, 2, 1, 1, 8, 10, 3, 9, 5, 6] | [ 4 3 7 11 10 9 2 5 6 8 1 6] |
| 12-717 | [0, 3, 7, 2, 1, 10, 8, 9, 5, 11, 4, 6] | [ 3 4 7 11 9 10 1 8 6 5 2 6] | 12-797 | [0, 4, 7, 3, 9, 2, 1, 11, 8, 10, 5, 6] | [ 4 3 8 6 5 11 10 9 2 7 1 6] |
| 12-718 | [0, 3, 7, 2, 1, 11, 10, 4, 9, 5, 6] | [ 3 4 7 11 10 2 1 6 9 4 5 6] | 12-798 | [0, 4, 7, 3, 10, 8, 9, 11, 5, 2, 1, 6] | [ 4 3 8 7 10 1 2 6 9 11 5 6] |
| 12-719 | [0, 3, 7, 4, 5, 10, 8, 2, 1, 9, 11, 6] | [ 3 4 9 1 5 10 6 11 8 2 7 6] | 12-799 | [0, 4, 7, 3, 1, 2, 8, 5, 10, 9, 11, 6] | [ 4 3 8 10 1 6 9 5 11 2 7 6] |

Table I. Listing of all the 918 AIS generators in prime form (continued).

| Label | Row | Intervals |
|---|---|---|
| 12-800 | [0, 4, 7, 5, 11, 8, 9, 2, 1, 3, 10, 6] | [ 4 3 10 6 9 1 5 11 2 7 8 6] |
| 12-801 | [0, 4, 7, 5, 11, 8, 1, 3, 2, 9, 10, 6] | [ 4 3 10 6 9 5 2 11 7 1 8 6] |
| 12-802L | [0, 4, 7, 5, 2, 1, 9, 3, 8, 10, 11, 6] | [ 4 3 10 9 11 8 6 5 2 1 7 6] |
| 12-803 | [0, 4, 9, 10, 7, 3, 5, 8, 2, 1, 11, 6] | [ 4 5 1 9 8 2 3 6 11 10 7 6] |
| 12-804 | [0, 4, 9, 10, 7, 5, 8, 2, 1, 3, 11, 6] | [ 4 5 1 9 10 3 6 11 2 8 7 6] |
| 12-805P | [0, 4, 9, 10, 8, 5, 11, 7, 2, 1, 3, 6] | [ 4 5 1 10 9 6 8 7 11 2 3 6] |
| 12-806 | [0, 4, 9, 10, 8, 7, 2, 11, 5, 1, 3, 6] | [ 4 5 1 10 11 7 9 6 8 2 3 6] |
| 12-807L | [0, 4, 8, 9, 11, 2, 8, 7, 3, 1, 10, 5, 6] | [ 4 5 2 3 6 11 8 10 9 7 1 6] |
| 12-808P | [0, 4, 9, 11, 2, 1, 7, 3, 10, 8, 5, 6] | [ 4 5 2 3 11 6 8 7 10 9 1 6] |
| 12-809L | [0, 4, 9, 11, 5, 1, 10, 8, 7, 2, 3, 6] | [ 4 5 2 6 8 9 10 11 7 1 3 6] |
| 12-810 | [0, 4, 9, 11, 5, 2, 3, 1, 8, 7, 10, 6] | [ 4 5 2 6 9 1 10 7 11 3 8 6] |
| 12-811 | [0, 4, 9, 11, 5, 7, 8, 2, 1, 10, 5, 3, 6] | [ 4 5 2 8 1 6 11 9 7 10 3 6] |
| 12-812 | [0, 4, 9, 11, 7, 1, 10, 8, 3, 2, 5, 6] | [ 4 5 2 8 6 9 10 7 11 3 1 6] |
| 12-813 | [0, 4, 9, 11, 8, 3, 2, 10, 1, 7, 5, 6] | [ 4 5 2 9 7 11 8 3 6 10 1 6] |
| 12-814 | [0, 4, 9, 11, 10, 7, 5, 1, 2, 8, 3, 6] | [ 4 5 2 11 9 10 8 1 6 7 3 6] |
| 12-815 | [0, 4, 9, 3, 11, 8, 5, 1, 7, 2, 3, 6] | [ 4 5 2 11 10 6 8 9 7 1 3 6] |
| 12-816 | [0, 4, 9, 3, 11, 1, 10, 8, 7, 2, 5, 6] | [ 4 5 6 8 2 9 10 11 7 3 1 6] |
| 12-817 | [0, 4, 9, 3, 11, 2, 1, 8, 10, 7, 5, 6] | [ 4 5 6 8 3 11 7 2 9 10 1 6] |
| 12-818S | [0, 4, 9, 7, 8, 11, 5, 2, 1, 3, 10, 6] | [ 4 5 10 1 3 6 9 11 2 7 8 6] |
| 12-819 | [0, 4, 9, 7, 8, 2, 11, 10, 5, 1, 3, 6] | [ 4 5 10 1 6 9 11 7 8 2 3 6] |
| 12-820 | [0, 4, 9, 7, 3, 5, 8, 2, 1, 10, 11, 6] | [ 4 5 10 8 2 3 6 11 9 1 7 6] |
| 12-821 | [0, 4, 9, 8, 10, 1, 7, 5, 2, 3, 11, 6] | [ 4 5 11 2 3 6 10 9 1 8 7 6] |
| 12-822 | [0, 4, 9, 8, 10, 7, 3, 1, 2, 5, 11, 6] | [ 4 5 11 2 9 8 10 1 3 6 7 6] |
| 12-823S | [0, 4, 9, 8, 11, 1, 7, 5, 2, 3, 10, 6] | [ 4 5 11 3 2 6 10 9 1 7 8 6] |
| 12-824S | [0, 4, 9, 8, 5, 7, 1, 11, 2, 3, 10, 6] | [ 4 5 11 9 2 6 10 3 1 7 8 6] |
| 12-825 | [0, 4, 10, 3, 5, 2, 1, 9, 7, 8, 11, 6] | [ 4 6 5 2 9 11 8 10 1 3 7 6] |
| 12-826 | [0, 4, 10, 3, 1, 9, 11, 8, 7, 2, 5, 6] | [ 4 6 5 10 8 2 9 11 7 3 1 6] |
| 12-827L | [0, 4, 10, 5, 8, 9, 7, 3, 2, 1, 11, 6] | [ 4 6 7 3 1 10 8 11 9 2 5 6] |
| 12-828L | [0, 4, 10, 5, 1, 11, 8, 7, 9, 2, 3, 6] | [ 4 6 7 8 10 9 11 2 5 1 3 6] |
| 12-829L | [0, 4, 10, 7, 9, 8, 1, 2, 5, 3, 11, 6] | [ 4 6 9 2 11 5 1 3 10 8 7 6] |
| 12-830 | [0, 4, 10, 7, 2, 1, 3, 8, 11, 9, 5, 6] | [ 4 6 9 7 11 2 5 3 10 8 1 6] |
| 12-831L | [0, 4, 10, 7, 3, 5, 8, 9, 2, 1, 11, 6] | [ 4 6 9 8 2 3 1 5 11 10 7 6] |
| 12-832L | [0, 4, 10, 7, 5, 8, 9, 2, 1, 3, 11, 6] | [ 4 6 9 10 3 1 5 11 2 8 7 6] |
| 12-833 | [0, 4, 10, 9, 11, 2, 7, 3, 1, 8, 5, 6] | [ 4 6 11 2 3 5 8 10 7 9 1 6] |
| 12-834 | [0, 4, 10, 9, 7, 3, 5, 8, 1, 2, 11, 6] | [ 4 6 11 10 8 2 3 5 1 9 7 6] |
| 12-835 | [0, 4, 11, 1, 3, 2, 8, 5, 10, 9, 7, 3, 6] | [ 4 7 2 1 6 9 5 11 10 8 3 6] |
| 12-836 | [0, 4, 11, 1, 9, 3, 2, 7, 10, 8, 5, 6] | [ 4 7 2 8 6 11 5 3 10 9 1 6] |
| 12-837 | [0, 4, 11, 1, 9, 7, 10, 3, 2, 8, 5, 6] | [ 4 7 2 8 10 3 5 11 6 9 1 6] |
| 12-838 | [0, 4, 11, 1, 10, 8, 7, 3, 9, 2, 5, 6] | [ 4 7 2 9 10 11 8 6 5 3 1 6] |
| 12-839 | [0, 4, 11, 2, 3, 9, 8, 10, 7, 5, 1, 6] | [ 4 7 3 1 6 11 2 9 10 8 5 6] |
| 12-840 | [0, 4, 11, 2, 8, 1, 10, 9, 7, 3, 5, 6] | [ 4 7 3 6 5 9 11 10 8 2 1 6] |
| 12-841S | [0, 4, 11, 2, 1, 3, 9, 7, 8, 5, 10, 6] | [ 4 7 3 11 2 6 10 1 9 5 8 6] |
| 12-842 | [0, 4, 11, 2, 1, 9, 3, 8, 10, 7, 5, 6] | [ 4 7 3 11 8 6 5 2 9 10 1 6] |
| 12-843 | [0, 4, 11, 5, 1, 10, 8, 7, 9, 2, 3, 6] | [ 4 7 6 8 9 10 11 2 5 1 3 6] |
| 12-844 | [0, 4, 11, 5, 2, 10, 9, 7, 8, 1, 3, 6] | [ 4 7 6 9 8 11 10 1 5 2 3 6] |
| 12-845L | [0, 4, 11, 5, 2, 1, 3, 8, 9, 7, 10, 6] | [ 4 7 6 9 11 2 5 1 10 3 8 6] |
| 12-846 | [0, 4, 11, 5, 2, 1, 9, 7, 8, 10, 3, 6] | [ 4 7 6 9 11 8 10 1 2 5 3 6] |
| 12-847 | [0, 4, 11, 7, 5, 2, 3, 9, 8, 10, 1, 6] | [ 4 7 8 10 9 1 6 11 2 3 5 6] |
| 12-848 | [0, 4, 11, 8, 10, 1, 9, 3, 2, 7, 5, 6] | [ 4 7 9 2 3 8 6 11 5 10 1 6] |
| 12-849 | [0, 4, 11, 8, 2, 1, 9, 7, 10, 3, 5, 6] | [ 4 7 9 6 11 8 10 3 5 2 1 6] |
| 12-850S | [0, 4, 11, 8, 7, 9, 3, 1, 2, 5, 10, 6] | [ 4 7 9 11 2 6 10 1 3 5 8 6] |
| 12-851 | [0, 4, 11, 9, 10, 7, 3, 5, 8, 2, 1, 6] | [ 4 7 10 1 9 8 2 3 6 11 5 6] |
| 12-852 | [0, 4, 11, 9, 5, 7, 8, 2, 1, 10, 3, 6] | [ 4 7 10 8 2 1 6 11 9 5 3 6] |
| 12-853 | [0, 4, 11, 9, 5, 10, 7, 8, 2, 1, 3, 6] | [ 4 7 10 8 5 9 1 6 11 2 3 6] |
| 12-854 | [0, 4, 11, 10, 8, 5, 7, 1, 9, 2, 3, 6] | [ 4 7 11 10 9 2 6 8 5 1 3 6] |
| 12-855 | [0, 4, 1, 2, 5, 10, 8, 7, 9, 3, 11, 6] | [ 4 9 1 3 5 10 11 2 6 8 7 6] |
| 12-856 | [0, 4, 1, 2, 7, 9, 5, 11, 10, 8, 3, 6] | [ 4 9 1 5 2 8 6 11 10 7 3 6] |
| 12-857 | [0, 4, 1, 2, 7, 5, 8, 10, 9, 3, 11, 6] | [ 4 9 1 5 10 3 2 11 6 8 7 6] |
| 12-858 | [0, 4, 1, 2, 9, 11, 5, 10, 8, 7, 3, 6] | [ 4 9 1 7 2 6 5 10 11 8 3 6] |
| 12-859S | [0, 4, 1, 2, 9, 11, 5, 3, 8, 7, 10, 6] | [ 4 9 1 7 2 6 10 5 11 3 8 6] |
| 12-860 | [0, 4, 1, 3, 8, 7, 2, 5, 11, 9, 10, 6] | [ 4 9 2 5 11 7 3 6 10 1 8 6] |
| 12-861L | [0, 4, 1, 3, 11, 9, 2, 8, 7, 10, 5, 6] | [ 4 9 2 8 10 5 6 11 3 7 1 6] |
| 12-862L | [0, 4, 1, 3, 11, 9, 8, 2, 7, 10, 5, 6] | [ 4 9 2 8 10 11 6 5 3 7 1 6] |
| 12-863 | [0, 4, 1, 3, 2, 7, 10, 8, 9, 5, 11, 6] | [ 4 9 2 11 5 3 10 1 8 6 7 6] |
| 12-864 | [0, 4, 1, 7, 3, 2, 5, 10, 8, 9, 11, 6] | [ 4 9 6 8 11 3 5 10 1 2 7 6] |
| 12-865 | [0, 4, 1, 7, 5, 8, 10, 9, 2, 3, 11, 6] | [ 4 9 6 10 3 2 11 5 1 8 7 6] |
| 12-866S | [0, 4, 1, 8, 5, 7, 3, 9, 8, 11, 5, 3, 2, 7, 10, 6] | [ 4 9 7 1 2 6 10 11 5 3 8 6] |
| 12-867P | [0, 4, 1, 8, 10, 9, 3, 11, 2, 7, 5, 6] | [ 4 9 7 2 11 6 8 3 5 10 1 6] |
| 12-868 | [0, 4, 1, 8, 11, 5, 3, 2, 7, 9, 10, 6] | [ 4 9 7 3 6 10 11 5 2 1 8 6] |
| 12-869 | [0, 4, 1, 8, 7, 9, 2, 10, 11, 5, 3, 6] | [ 4 9 7 11 2 5 8 1 6 10 3 6] |
| 12-870L | [0, 4, 1, 9, 11, 5, 10, 8, 7, 2, 3, 6] | [ 4 9 8 2 6 5 10 11 7 1 3 6] |
| 12-871 | [0, 4, 1, 9, 3, 2, 7, 5, 8, 10, 11, 6] | [ 4 9 8 6 11 5 10 3 2 1 7 6] |
| 12-872 | [0, 4, 1, 9, 7, 8, 10, 3, 2, 5, 11, 6] | [ 4 9 8 10 1 2 5 11 3 6 7 6] |
| 12-873L | [0, 4, 1, 11, 2, 7, 3, 9, 8, 10, 5, 6] | [ 4 9 10 3 5 8 6 11 2 7 1 6] |
| 12-874 | [0, 4, 1, 11, 7, 10, 9, 2, 8, 3, 5, 6] | [ 4 9 10 8 3 11 5 6 7 2 1 6] |
| 12-875 | [0, 4, 2, 3, 5, 10, 7, 1, 9, 8, 11, 6] | [ 4 10 1 2 5 9 6 8 11 3 7 6] |
| 12-876 | [0, 4, 2, 3, 10, 7, 9, 8, 11, 5, 1, 6] | [ 4 10 1 7 9 2 11 3 6 8 5 6] |
| 12-877 | [0, 4, 2, 5, 7, 8, 1, 10, 9, 3, 11, 6] | [ 4 10 3 2 1 5 9 11 6 8 7 6] |
| 12-878 | [0, 4, 2, 5, 7, 3, 9, 8, 1, 11, 10, 6] | [ 4 10 3 2 8 6 11 5 9 1 7 6] |
| 12-879 | [0, 4, 2, 7, 9, 3, 11, 10, 1, 8, 5, 6] | [ 4 10 5 2 6 8 11 3 7 9 1 6] |
| 12-880P | [0, 4, 2, 7, 10, 9, 3, 11, 1, 8, 5, 6] | [ 4 10 5 3 11 6 8 2 7 9 1 6] |
| 12-881L | [0, 4, 2, 7, 3, 9, 8, 11, 1, 10, 5, 6] | [ 4 10 5 8 6 11 3 2 9 7 1 6] |
| 12-882 | [0, 4, 2, 1, 7, 11, 3, 2 6, 10, 5, 9, 6] | [ 4 10 7 2 11 3 2 6 5 9 1 6] |
| 12-883 | [0, 4, 2, 9, 5, 11, 10, 7, 8, 1, 3, 6] | [ 4 10 7 8 6 11 9 1 5 2 3 6] |
| 12-884 | [0, 4, 2, 9, 8, 11, 7, 1, 10, 3, 5, 6] | [ 4 10 7 11 3 8 6 9 5 2 1 6] |
| 12-885 | [0, 4, 2, 9, 8, 5, 10, 11, 1, 7, 3, 6] | [ 4 10 7 11 9 5 1 2 6 8 3 6] |
| 12-886 | [0, 4, 2, 11, 1, 8, 7, 10, 3, 9, 5, 6] | [ 4 10 9 2 7 11 3 5 6 8 1 6] |
| 12-887 | [0, 4, 2, 11, 1, 9, 3, 8, 7, 10, 5, 6] | [ 4 10 9 2 8 6 5 11 3 7 1 6] |
| 12-888 | [0, 4, 2, 11, 5, 1, 8, 7, 9, 10, 3, 6] | [ 4 10 9 6 8 7 11 2 1 5 3 6] |
| 12-889 | [0, 4, 2, 1, 3, 9, 5, 10, 7, 8, 11, 6] | [ 4 10 11 2 6 8 5 9 1 3 7 6] |
| 12-890 | [0, 4, 2, 1, 8, 9, 3, 11, 5, 10, 7, 3, 6] | [ 4 10 11 7 1 2 6 5 9 8 3 6] |
| 12-891P | [0, 4, 2, 1, 8, 5, 11, 7, 9, 10, 3, 6] | [ 4 10 11 7 9 6 8 2 1 5 3 6] |
| 12-892 | [0, 4, 2, 1, 9, 11, 5, 10, 7, 8, 3, 6] | [ 4 10 11 8 2 6 5 9 1 7 3 6] |
| 12-893L | [0, 4, 2, 1, 9, 3, 8, 5, 7, 10, 11, 6] | [ 4 10 11 8 6 5 9 2 3 1 7 6] |
| 12-894P | [0, 4, 2, 1, 1, 10, 9, 5, 7, 8, 11, 6] | [ 4 10 11 9 5 6 8 2 1 3 7 6] |
| 12-895 | [0, 4, 3, 5, 8, 9, 2, 10, 7, 1, 11, 6] | [ 4 11 2 3 1 5 8 9 6 10 7 6] |
| 12-896S | [0, 4, 3, 5, 8, 1, 7, 2, 11, 9, 10, 6] | [ 4 11 2 3 5 6 7 9 10 1 8 6] |
| 12-897 | [0, 4, 3, 5, 10, 1, 9, 7, 8, 2, 11, 6] | [ 4 11 2 5 3 8 10 1 6 9 7 6] |
| 12-898 | [0, 4, 3, 5, 1, 5, 2, 9, 7, 2, 10, 1, 6] | [ 4 11 2 6 9 1 10 7 8 3 5 6] |
| 12-899S | [0, 4, 3, 5, 2, 7, 1, 8, 11, 9, 10, 6] | [ 4 11 2 9 5 6 7 3 10 1 8 6] |
| 12-900P | [0, 4, 3, 5, 2, 7, 1, 9, 10, 8, 11, 6] | [ 4 11 2 9 5 6 8 1 10 3 7 6] |
| 12-901L | [0, 4, 3, 5, 2, 8, 1, 9, 7, 10, 11, 6] | [ 4 11 2 9 6 5 8 10 3 1 7 6] |
| 12-902 | [0, 4, 3, 5, 2, 9, 10, 8, 11, 7, 1, 6] | [ 4 11 2 9 7 1 10 3 8 6 5 6] |
| 12-903 | [0, 4, 3, 8, 10, 1, 7, 2, 11, 9, 5, 6] | [ 4 11 5 2 3 6 7 9 10 8 1 6] |
| 12-904 | [0, 4, 3, 8, 10, 7, 1, 11, 2, 9, 5, 6] | [ 4 11 5 2 9 6 10 3 7 8 1 6] |
| 12-905S | [0, 4, 3, 8, 11, 1, 1, 7, 5, 2, 9, 10, 6] | [ 4 11 5 3 2 6 10 9 7 1 8 6] |
| 12-906 | [0, 4, 3, 8, 2, 11, 1, 9, 7, 10, 5, 6] | [ 4 11 5 6 9 2 8 10 3 7 1 6] |
| 12-907S | [0, 4, 3, 8, 5, 7, 1, 11, 2, 9, 10, 6] | [ 4 11 5 9 2 6 10 3 7 1 8 6] |
| 12-908 | [0, 4, 3, 9, 5, 7, 10, 8, 1, 2, 11, 6] | [ 4 11 6 8 2 3 10 5 1 9 7 6] |
| 12-909 | [0, 4, 3, 9, 5, 2, 7, 8, 10, 1, 11, 6] | [ 4 11 6 8 9 5 1 2 3 10 7 6] |
| 12-910 | [0, 4, 3, 10, 1, 9, 11, 8, 2, 7, 5, 6] | [ 4 11 7 3 8 2 9 6 5 10 1 6] |
| 12-911 | [0, 4, 3, 10, 1, 11, 7, 9, 2, 8, 5, 6] | [ 4 11 7 3 10 8 2 5 6 9 1 6] |
| 12-912 | [0, 4, 3, 10, 1, 11, 8, 2, 7, 9, 5, 6] | [ 4 11 7 3 10 9 6 5 2 8 1 6] |
| 12-913 | [0, 4, 3, 11, 2, 9, 10, 8, 5, 7, 1, 6] | [ 4 11 8 3 7 1 10 9 2 6 5 6] |
| 12-914 | [0, 4, 3, 11, 5, 2, 9, 7, 8, 10, 1, 6] | [ 4 11 8 6 9 7 10 1 2 3 5 6] |
| 12-915S | [0, 4, 3, 1, 8, 11, 5, 2, 7, 9, 10, 6] | [ 4 11 10 7 3 6 9 5 2 1 8 6] |
| 12-916 | [0, 4, 3, 1, 9, 11, 8, 2, 7, 10, 5, 6] | [ 4 11 10 8 2 9 6 5 3 7 1 6] |
| 12-917L | [0, 4, 3, 1, 9, 2, 8, 5, 7, 10, 11, 6] | [ 4 11 10 8 5 6 9 2 3 1 7 6] |

Table I. Listing of all the 918 AIS generators in prime form (continued).

**Topology of the AIS space**. In order to map what Morris and Starr call "chains of relations" among AIS, we leverage the approach outlined in (Buongiorno Nardelli, Topology of Networks in Generalized Musical Spaces 2020), where the author introduces the concept of network representation of generalized musical spaces.[2] Network analysis methods exploit the use of graphs or networks as convenient tools for modeling relations in large data sets. If the elements of a data set are thought of as "nodes", then the emergence of pairwise relations between them, "edges", yields a network representation of the underlying set. Similarly to social networks, biological networks and other well-known real-world complex networks, entire dataset of musical structures can be treated as networks, where each individual musical entity, in this case the prime form AIS, is represented by a node, and a pair of nodes is connected by a link if the respective two objects exhibit a certain level of similarity according to a specified quantitative metric. Pairwise similarity relations between nodes are thus defined through the introduction of a measure of "distance" in the network (Albert and Barabási 2002). In this work we use a metric based on the Euclidean parsimonious voice leading distance between two AIS. Given two AIS, A=[$a_0,a_1,a_2,a_3,a_4,a_5,a_6,a_7,a_8,a_9,a_{10},a_{11}$] and B=[$b_0,b_1,b_2,b_3,b_4,b_5,b_6,b_7,b_8,b_9,b_{10},b_{11}$], we define the parsimonious voice-leading distance $d$ as

$$d = \sqrt{\sum_i \min\left(|a_i - b_i|^2, \left||a_i - b_i| - 12\right|^2\right)}.$$

Using the above metric, we build a network where the individual nodes are the AIS in prime form, and links are drawn if the distance between the two nodes is smaller than a threshold. Using this procedure, we obtain the network of Figure 1.

---

[2] This approach has also been recently used to derive a framework for representation and analysis of tonal music (Buongiorno Nardelli, Tonal harmony, the topology of dynamical score networks and the Cinese postman problem 2020).

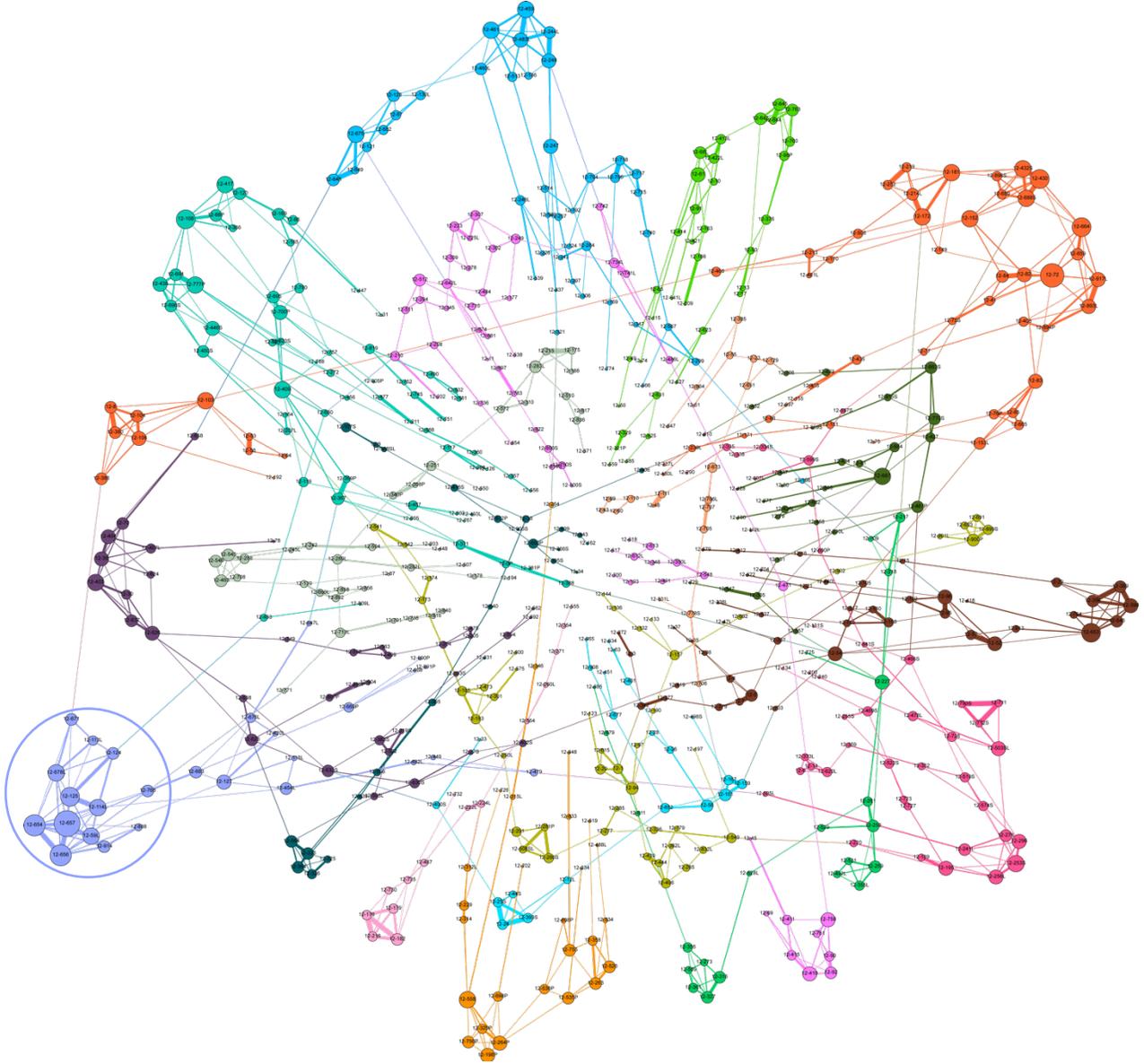

Figure 1. Giant connected component of the network representation of the 918 AIS in prime form with a threshold $d^2 \leq 20$. In the full network, the farthest distance between two AIS is $d^2 = 306$, so only a small fraction of all the links is displayed here. Network drawing and analysis are done using the Gephi app at www.gephi.org. A high-resolution figure and the gephi input file are available on-line as supplemental information to this paper.

Given a network we can perform many statistical operations that shed light on the internal structure of the data. In this work we will consider only two of such measures, degree centrality and modularity class. (Barabasi and Posfai 2016) The degree of a node is measured by the number of edges that depart from it. It is a local measure of the relative "importance" of a node in the network. Modularity is a measure of the strength of division of a network into communities: high modularity (above 0.6 in a scale from 0 to 1) corresponds to networks that have a clearly visible community structure. (Zinoviev 2018). Isolating communities through modularity measures provides a way to operate within regions of higher similarity, and thus, provide the "re-ordering scheme" invoked by Morris and Starr.

In Figure 1 we display the giant connected component (graph in which we show only vertices that are connected to each other by paths, 648 of the total 918 nodes) of the AIS network constructed with a distance threshold of $d^2 \leq 20$: there is a link between all the nodes whose relative distance is less than the threshold, and the link thickness is inversely proportional to the distance (closer nodes are connected by thicker bonds). It should be noted that with the chosen threshold there are also 111 AIS that form a subset of completely unconnected nodes of zero degree. Node radii are proportional to their degree, and nodes that display higher degree of similarity are grouped in modularity classes where all the nodes of the same community have the same color. From a simple inspection of Figure 1, it is evident the highly modular community structure of the AIS network: the average degree is ~ 3 and the modularity is 0.9. The network of Figure 1 is undirected: links do not possess a preferential direction.

The first observation of particular interest is that the AIS network exhibits some scale-free properties, that is, the distribution of node degrees follows a power law (Barabasi and Albert 1999). In Figure 2 we show the plot of the degree distribution of the AIS network, where the data are fitted to a modified power law distribution of the form $p(x) \propto x^{-\alpha}e^{-\lambda x}$. As we can see, the scaling of the distribution is very well represented by such functional form. This implies that the network is characterized by a few nodes with many connections (hubs) while the remainder exhibits relatively few links. (Barabasi and Posfai 2016) As we will see in the following, this is a property that provides directly "interesting compositional devices for pitch-ordering" (Morris and Starr, The Structure of All-Interval Series 1974).[3]

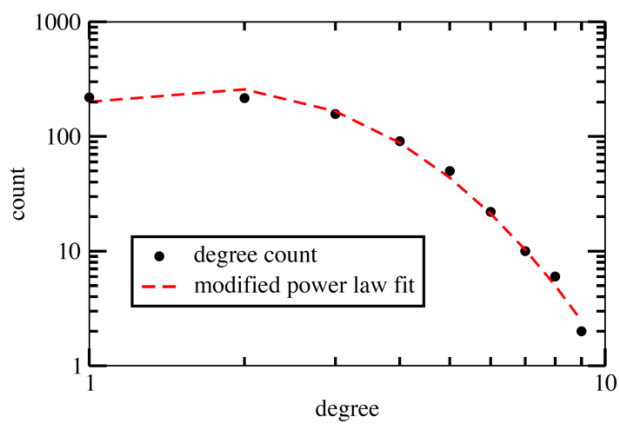

Figure 2. Loglog plot of the degree distribution in the AIS network. Dashed line is the best fit to a modified power law distribution (see text), readily demonstrating that indeed the degree distribution in the network displays some scale-free property.

To get a better sense of the network topology and of the internal relationships between nodes, let's focus on the group of nodes circled in blue in the lower left corner of the graph in Figure 1 and analyze their mutual links. Here we find a tightly connected cluster of AIS, centered around 12-

---

[3] The computational tools for the generation and analysis of network of generalized musical spaces can be found in the MUSICNTWRK software package (Buongiorno Nardelli, MUSICNTWRK: data tools for music theory, analysis and composition, 2019).

657, that is clearly a hub for this community. Moreover, 12-657, 12-656, 12-654, 12-125, 12-114L and 12-59L form a "clique", a subset of vertices of an undirected graph such that every two distinct vertices in the clique are adjacent. The rows of the clique, their interval sequence, and the square of the parsimonious voice-leading distance are summarized in Example 4. The first observation is that the square of the distance is always an even number. This is the reflection of the fact that we cannot change only one pitch in the row without breaking the all-interval property and was already noted by Morris and Starr as "swapping relationship". Indeed, a close inspection at the rows that exhibit a $d^2 = 2$ shows that this is the case: 12-657 is transformed into 12-656 by the swapping of 2 pc (1 ↔ 2). This re-ordering corresponds to an analogous swapping in the interval sequence: in the new sequence the intervals of 9 and 8 semitones exchange places, as do the ones of 4 and 5. For $d^2 = 4$, such as for 12-657 and 12-654, the swapping involves 4 pc separated by a semitone (11 ↔ 10 and 2 ↔ 1). Naturally, distances become larger if more pc are swapped and also if the swapping involves pc that are not adjacent, for instance, between 12-657 and 12-59L there are four swapping of one semitone, and one of one tone. The network of Figure 1 is the epitome of the re-ordering scheme invoked by Morris and Starr and provides a navigational map that allows composers and theorists to explore the broad range of relations among different AIS classes. Moreover, since the parsimonious voice-leading distance is invariant under the operations of inversion (I) and retrograde (R), the network of Figure 1, obtained for the prime forms P of the AIS, is isomorphic to the networks of I(P) and R(P), the internal relationships being the same. This invariance is lost under the two other operations, and M(P) and Q(P) will generate two independent networks.

| Label | Row | Intervals |
|---|---|---|
| 12-657 | [0, 3, 4, 11, 5, 2, 10, 8, 7, 9, 1, 6] | [ 3 1 7 6 9 8 10 11 2 4 5 6] |
| 12-656 | [0, 3, 4, 11, 5, 1, 10, 8, 7, 9, 2, 6] | [ 3 1 7 6 8 9 10 11 2 5 4 6] |
| 12-654 | [0, 3, 4, 10, 5, 1, 11, 8, 7, 9, 2, 6] | [ 3 1 6 7 8 10 9 11 2 5 4 6] |
| 12-125 | [0, 1, 5, 11, 4, 2, 10, 7, 9, 8, 3, 6] | [ 1 4 6 5 10 8 9 2 11 7 3 6] |
| 12-114L | [0, 1, 5, 10, 4, 2, 11, 7, 9, 8, 3, 6] | [ 1 4 5 6 10 9 8 2 11 7 3 6] |
| 12-59L | [0, 1, 4, 10, 5, 3, 11, 8, 7, 9, 2, 6] | [ 1 3 6 7 10 8 9 11 2 5 4 6] |

|  | 12-657 | 12-656 | 12-654 | 12-125 | 12-114L | 12-59L |
|---|---|---|---|---|---|---|
| **12-657** | 0 | 2 | 4 | 16 | 18 | 8 |
| **12-656** | 2 | 0 | 2 | 14 | 16 | 10 |
| **12-654** | 4 | 2 | 0 | 16 | 14 | 8 |
| **12-125** | 16 | 14 | 16 | 0 | 2 | 12 |
| **12-114L** | 18 | 16 | 14 | 2 | 0 | 10 |
| **12-59L** | 8 | 10 | 8 | 12 | 10 | 0 |

Example 4. Top: the rows of the clique and their interval sequence; bottom: square of the voice-leading parsimonious distance between AIS.

To classify the re-ordering relations between AIS in the spirit of Morris and Starr, we can further order the prime-form rows in terms of their swapping properties: there is a limited number of AIS that allow for a single swap of a given distance. As an illustration, we can look for the AIS that allow a single swap ($d = \sqrt{2}$): there are only 42 close-coupled AIS in prime form that enjoy this property.[4] As we mentioned above, there is also a subset of AIS that do not show any connection with any other (degree=0) within our choice of threshold for link generation. For these AIS, that we name "hermit", we do not find simple swapping relations that would transform them into other AIS in prime form.[5] Of course, this classification is subject to the choice of threshold, but with the help of the computational tools developed for this paper and distributed as supplemental material, the readers should be able to design any relational framework for their use.

**Conclusions**. In this work we have taken up the challenge posed by Morris and Starr in their seminal paper on the generation of all-interval series, and derive a re-ordering scheme based on network representation that links all AISs together by chains of relations, thus designing the elegant description of the generation of AISs envisioned by the authors, and providing theorists and composers with the computational tools needed to design interesting compositional devices for pitch-ordering.

**Acknowledgements**. I owe this paper to my friend and colleague David Bard-Schwarz, who challenged me to tackle the mapping of the AIS once and for all and has been a staunch supporter of this project since its inception. I am also particularly grateful for the support of the University of North Texas through the Center for Experimental Music and Intermedia (CEMI) and the Initiative for Advanced Research in Technology and the Arts (iARTA).

**Marco Buongiorno Nardelli** is University Distinguished Research Professor in the Department of Physics and the Division of Composition at the University of North Texas.

**Supplemental online material**. Supplemental on-line material (high resolution Figure 1, gephi input file for manual exploration of the network, and the IPython notebook with the code used to generate all the results) are available at https://github.com/marcobn/TheHichhikersGuideToAIS.

---

[4] These are the 42 close coupled AIS: (12-5 - 12-7), (12-8 - 12-380), (12-13 - 12-17), (12-18 - 12-19), (12-24 - 12-25S), (12-24 - 12-393S), (12-48 - 12-111), (12-52 - 12-57), (12-53 - 12-58), (12-82 - 12-84), (12-95 - 12-96), (12-114L - 12-125), (12-159 - 12-167), (12-172 - 12-214L), (12-175 - 12-215), (12-176 - 12-182), (12-176 - 12-216), (12-219 - 12-252), (12-244L - 12-248), (12-367 - 12-369P), (12-389 - 12-390), (12-389 - 12-394), (12-430 - 12-432S), (12-430 - 12-688S), (12-459 - 12-483L), (12-462 - 12-485), (12-644 - 12-645), (12-648 - 12-649), (12-654 - 12-656), (12-656 - 12-657), (12-694 - 12-777P), (12-701 - 12-782), (12-707 - 12-786L), (12-711 - 12-712S), (12-711 - 12-793S), (12-718 - 12-796), (12-723 - 12-727), (12-734L - 12-741L), (12-753 - 12-754), (12-821 - 12-823S), (12-828L - 12-843), (12-893L - 12-917L)

[5] The 111 hermit AIS are: 12-22, 12-32, 12-46L, 12-62, 12-118, 12-135, 12-138, 12-140, 12-154L, 12-168, 12-184, 12-201, 12-232, 12-235, 12-236, 12-274L, 12-275, 12-279, 12-298, 12-311L, 12-330, 12-349, 12-370L, 12-374, 12-378, 12-383, 12-384, 12-391, 12-392, 12-397, 12-398, 12-399, 12-425, 12-429, 12-433, 12-442, 12-455, 12-465, 12-470, 12-476L, 12-495L, 12-497L, 12-527L, 12-533, 12-543, 12-544, 12-560, 12-561, 12-568, 12-582, 12-601, 12-608, 12-616, 12-621, 12-635, 12-636, 12-655, 12-661, 12-685, 12-716, 12-721, 12-733, 12-737, 12-743, 12-759, 12-762S, 12-772, 12-773, 12-775, 12-785, 12-787, 12-789, 12-797, 12-804, 12-807L, 12-811, 12-812, 12-813, 12-814, 12-824S, 12-825, 12-829L, 12-830, 12-838, 12-839, 12-845L, 12-849, 12-851, 12-852, 12-853, 12-855, 12-857, 12-860, 12-861L, 12-863, 12-868, 12-869, 12-870L, 12-871, 12-872, 12-875, 12-878, 12-881L, 12-883, 12-889, 12-890, 12-898, 12-903, 12-907S, 12-909, 12-912.